\documentclass[article]{aa}  
\usepackage{lineno}
\usepackage{graphicx}
\usepackage{txfonts}
\begin{document}

   \title{Optical spectroscopy of blazars for the Cherenkov Telescope Array \thanks{Based on observations collected at the European Organisation for Astronomical Research in the Southern Hemisphere, Chile, under programmes P103.B-0430(A). The raw FITS data files are available in the ESO archive. Some of the data presented herein were obtained at the W. M. Keck Observatory, which is operated as a scientific partnership among the California Institute of Technology, the University of California, and the National Aeronautics and Space Administration. The Observatory was made possible by the generous financial support of the W. M. Keck Foundation. Based on observations made with the Southern African Large Telescope (SALT) under programme 2019-2-SCI-044 (PI E. Kasai)} }

 \author{P.~Goldoni\inst{1}
  \and S. Pita\inst{1}
  \and C. Boisson\inst{2}
  \and W. Max-Moerbeck\inst{3}
  \and E. Kasai\inst{4}
  \and D. A. Williams\inst{5}
  \and F. D'Ammando\inst{6}
  \and V. Navarro-Aranguiz\inst{3}
  \and M. Backes\inst{4,7}
  \and U. Barres de Almeida\inst{8}
  \and J. Becerra-Gonzalez\inst{9,10}
  \and G. Cotter\inst{11}
  \and O. Hervet\inst{5}
  \and J.-P. Lenain\inst{12}
  \and E. Lindfors\inst{13}
  \and H. Sol\inst{2}
  \and S. Wagner\inst{14}
}

   \institute{APC, AstroParticule et Cosmologie, Universit\'{e} Paris Diderot, CNRS/IN2P3, CEA/Irfu, Observatoire de Paris, Sorbonne Paris Cit\'{e}, 10, rue Alice Domon et L\'{e}onie Duquet, F-75006 Paris, France 
\and
LUTH, Observatoire de Paris,  PSL Research University, CNRS,Universit\'{e} Paris Diderot, Meudon, France  
\and
Departamento de Astronomía, Universidad de Chile, Camino El Observatorio 1515, Las Condes, Santiago, Chile
\and
Department of Physics, University of Namibia, Private Bag 13301, Windhoek, Namibia
\and
Santa Cruz Institute for Particle Physics and Department of Physics, University of California, Santa Cruz, Santa Cruz, CA
\and
INAF - Istituto di Radioastronomia, Via Gobetti 101, I-40129 Bologna, Italy
\and
Centre for Space Research, North-West University, Potchefstroom 2520, South Africa
\and
Centro Brasileiro de Pesquisas Físicas (CBPF), Rua Dr. Xavier Sigaud 150, 22290-180 Rio de Janeiro, Brazil
\and 
Universidad de La Laguna (ULL), Departamento de Astrof\'isica, E-38206 La Laguna, Tenerife, Spain
\and
Instituto de Astrof\'isica de Canarias (IAC), E-38200 La Laguna, Tenerife, Spain
\and
University of Oxford, Oxford Astrophysics, Denys Wilkinson Building, Keble Road, Oxford, OX1 3RH, United Kingdom
\and
Sorbonne Universit\'{e}, Universit\'{e} Paris Diderot, Sorbonne Paris Cit\'{e}, CNRS/IN2P3, Laboratoire de Physique Nucl\'{e}aire et de Hautes Energies, LPNHE, 4 Place Jussieu, F-75252 Paris, France
\and
Finnish Centre for Astronomy with ESO (FINCA), Quantum, Vesilinnantie 5, FI-20014, University of Turku, Finland
\and
Landessternwarte, Universit\"at Heidelberg, K\"onigstuhl 12, D 69117 Heidelberg, Germany
}

   \date{}

  \abstract
   { Blazars are the most numerous  class of high-energy (HE; E $\sim$ 50 MeV -- 100 GeV) and very high-energy (VHE; E $\sim$ 100 GeV -- 10 TeV) gamma-ray emitters. Currently, a measured spectroscopic redshift is available for only about 50\% of gamma-ray BL Lacertae objects (BL Lacs), mainly due to the difficulty in measuring reliable redshifts from their nearly featureless continuum-dominated optical spectra. The knowledge of the redshift is fundamental for understanding the emission from blazars, for population studies and also for indirect studies of the extragalactic background light and searches for Lorentz invariance violation and axion-like particles using blazars.}
   {This paper is the first in a series of papers that aim to measure the redshift of a sample of blazars likely to be detected with the upcoming Cherenkov Telescope Array (CTA), a ground-based gamma-ray observatory.}
   {Monte Carlo simulations were performed to select those hard spectrum gamma-ray blazars detected with the Fermi-LAT telescope still lacking redshift measurements, but likely to be detected by CTA in 30 hours of observing time or less. Optical observing campaigns involving deep imaging and spectroscopic observations were organised to efficiently constrain their redshifts. We performed deep medium- to high-resolution  spectroscopy of 19 blazar optical counterparts with the Keck II, SALT, and  ESO NTT telescopes. We searched systematically for spectral features and, when possible, we estimated the contribution of the host galaxy to the total flux.}
   {We measured eleven firm spectroscopic redshifts with values ranging from 0.1116 to 0.482, one tentative redshift, three redshift lower limits including
one at z $\ge$ 0.449 and another at z $\ge$ 0.868. Four 
BL Lacs  show featureless spectra.}
  {}


   \maketitle
%

\section{Introduction}

Blazars, a subclass of radio-loud active galactic nuclei (AGN),  display the most extreme observational properties of all types of AGN. They show unique characteristics such as beamed non-thermal emission from the radio to gamma-rays, strong ($\ge$ 3\%) polarised optical \citep[see e.g.][]{Angel80,Angelakis16} and radio \citep{Lis11} emission, variability from a few percent up to a few orders of magnitude on different timescales at all wavelengths \citep[see e.g.][]{Wag95,Falo14}, and for some, on the Very Long Baseline Interferometry scale,  the presence of superluminal radio blobs \citep[see e.g.][]{VerCo94}. These characteristics are generally explained by strong Doppler amplification of the jet emission with Lorentz factors of up to $\sim$40 \citep[e.g.][]{jorstad17}, the latter being seen at a very small angle ($\theta \le 5 \deg $) with respect to the jet axis.

Blazars are classified into flat-spectrum radio quasars (FSRQs) and BL Lacertae objects (BL Lacs), an important distinctive feature being the presence of 
broad (a few thousand km/s) and luminous ($\ge$ 10$^{42}$ erg s$^{-1}$) emission lines in the optical spectra of the former and their weakness or even absence in the latter. The strong emission lines in FSRQs are produced by ionised gas in the vicinity of the black hole. Their presence may be an indication that the accretion regime in FSRQs is different with respect to that in BL Lacs \citep[e.g.][]{Ghis11}. Both blazar subtypes present two broad distinct components in their spectral energy distribution (SED), the low-energy component peaking in the infrared to X-ray band and the high-energy component peaking in the MeV to TeV band. The lower energy component is due to synchrotron radiation, while the higher energy one is generally ascribed to inverse Compton upscattering of lower energy photons off the population of accelerating electrons in the jet, though a  hadronic component may also be responsible for the second SED peak \citep[see e.g.][]{Muc01,Cer15}. The location of the low-energy peak can be used to subdivide BL Lacs into four different classes \citep[see e.g.][]{Pado95,Cost01}. If the peak is at a frequency lower than 10$^{14}$ Hz, the BL Lac object is a low-frequency peaked BL Lac object (LBL); if it is between 10$^{14}$ and 10$^{15}$ Hz it is an intermediate-frequency peaked BL Lac object (IBL); if it is between 10$^{15}$ and 10$^{17}$ Hz it is a high-frequency peaked BL Lac object (HBL); and finally if it is at a frequency higher than 10$^{17}$ Hz, it is an extreme high-frequency peaked BL Lac object (EHBL). A slightly different classification scheme, defined in \citet{Abdo10}, is used in the Third Fermi High Energy Catalog (3FHL) \citep{Fer3FHL17} where both FSRQs and BL Lacs are divided into low-synchrotron peaked (LSP, equivalent to LBL for BL Lacs), intermediate-synchrotron peaked (ISP, equivalent to IBL for BL Lacs), and high-synchrotron peaked (HSP, equivalent to HBL and EHBL for BL Lacs).

In the HE regime, covered since 2008 by the Large Area Telescope (LAT) on board the {\em Fermi} satellite, blazars account for more than 3400  of $\sim$5000 ($\sim$
68\%) detected sources, as reported in the Data Release 2 of the Fourth Fermi Large Area Telescope  catalogue of Gamma-Ray sources (4FGL-DR2) \citep{Bal20}; 730 of them are FSRQs, 1190 are BL Lacs, and 1517 are blazar candidates of uncertain types (BCUs). In the VHE regime  the current generation of Imaging Atmospheric Cherenkov Telescopes (IACTs) (H.E.S.S.\footnote{https://www.mpi-hd.mpg.de/hfm/HESS/}, MAGIC\footnote{https://magic.mpp.mpg.de}, and VERITAS\footnote{https://veritas.sao.arizona.edu}), has detected 75 blazars, about one-half of the total number of identified sources\footnote{http://tevcat.uchicago.edu}. Most of these blazars, 64 out of 75, are BL Lacs. Of these 64 TeV BL Lacs, we note that 13 still lack spectroscopic redshift values.

In the next few years a new facility, the Cherenkov Telescope Array (CTA\footnote{https://www.cta-observatory.org}) will become operational with a northern site in the Canary Islands (Spain) and a southern site in the Atacama desert (Chile). With a 20 GeV - 300 TeV energy range and a sensitivity approximately ten times better than the current generation of IACTs, it is expected to detect hundreds of blazars according to current estimates, thus opening the possibility of population studies with a significant sample size  \citep{Sctashort19}. 

  The CTA will enable substantial progress on gamma-ray population studies by deepening existing surveys, and will permit more detailed studies of the VHE gamma-ray emission and its origin  \citep{Sol13}. Furthermore, the stellar optical and near-infrared (NIR) radiation, called the extragalactic background light (EBL) \citep[see e.g.] [] {Hau01,Bit15,abda17,fermi18,abey19,acc19}, acts as a source of opacity for the gamma rays from blazars. This effect can be used to derive within the same fitting process the properties of the EBL from its imprint on the VHE spectrum of blazars and the intrinsic VHE spectra of these sources \citep[see e.g.][]{Dom15}. The pro\-pa\-gation of VHE gamma-ray radiation can be used to investigate topics related to cosmology or fundamental physics: the properties of the intergalactic magnetic field (IGMF)  \citep[see e.g.][]{Acke18,Aha94, Alves19}, the possible existence of axion-like particles \citep[see e.g.][]{Miriz07,Deang11,Abra13}, an independent measurement of the Hubble constant H0 \citep{Sala94}, or the search for Lorentz Invariance Violation \citep{Kifu99}. More details on these subjects can be found in \citet[][ and references therein]{Sctashort19}. 
 
  The capabilities of CTA to advance these subjects are presented in \citet{CTAcons20}. Finally, another important science case for the measurement of the redshifts of BL Lacs is the first evidence of neutrino emission from these sources  \citep{IceCube18a,IceCube18b,Fra20,Giom20,Pal20}. A precise estimation of the total luminosity is necessary to fully understand the role of hadrons in the jet, which implies the knowledge of the redshift \citep[see e.g.][]{Pai18}.
 
 For BL Lacs these exciting possibilities are hampered by the difficulty in measuring reliable redshifts from their nearly featureless, continuum-dominated optical spectra.  In optical spectroscopy one of the distinctive
properties of BL Lacs  is that they are objects with weak emission lines. The limit is usually set at an equivalent width (EW) of 5 \AA~\citep{urr95}, but it is known that the emission lines of BL Lacs can sometimes be brighter than that \citep[see e.g.] []{Stick91}. High signal-to-noise (S/N) spectra are usually needed to detect these weak lines.
 For this reason the measurement of the redshifts of BL Lacs is a challenging task; spectroscopic observations are often unsuccessful and a large fraction of BL Lacs lack redshifts. Recognising this, several spectroscopic campaigns to measure the redshifts of gamma-ray BL Lacs detected by {\em Fermi}-LAT have been organised. An early extensive effort was made   by \citet{Shaw13}, who reported on rather deep observations from the southern and northern sites of 372 BL Lacs. Adding previous literature results, they obtained a sample with 44\% redshift completeness whose median redshift is $z_{\rm med}$ = 0.33. Since then new {\em Fermi}-LAT source catalogues have been published, the latest being the 4FGL-DR2 \citep{Bal20}, and new associations with blazars have been produced  \citep[see e.g.] [] {Ace13,Arsi15,Arsi17,Kau19}. Several groups have performed spectroscopic campaigns often focused on these new {\em Fermi}-LAT BL Lacs and BCUs. One group \citep[see e.g.] [] {Pai17a,Lan18,Pai20} has pursued high S/N observations mainly from the Gran Telescopio Canarias (GTC) of different subsamples of gamma-ray blazars selected for being unidentified or for being likely to be detected at  very high energies
from simple estimations.  A very extensive series of papers \citep[see e.g.] []{Mas13,Pag14,Mas15a,Mas15b} has pursued the identification and redshift measurement of {\em Fermi}-LAT BCUs with low- and medium-sensitivity observations from both hemispheres. Other less extensive campaigns with similar aims, which have targeted smaller samples, include those of \citet{Mase13}, \citet{Marc16}, and \citet{Klin17}. A synthesis of these recent observations has been presented by \citet{Pena20} for a total of 416 BL Lacs and BCUs, 311 taken from their own observations and 105 from recent literature including the ones cited above. Only about 30\% of these objects have spectroscopic redshift values,  and their median redshift is $z_{\rm med}$ = 0.285. 

This incompleteness in redshift determination implies that it is very difficult to determine the properties of blazars as a population. A fundamental quantity such as luminosity is not determined for more than half of them. As a consequence, the blazar sequence (i.e. the observation  that the peak frequency of the blazar SED becomes redder with increasing peak luminosity)  is still a very controversial subject.
 It has been interpreted as being due to differences in radiative cooling among blazar classes \citep{Ghis17} or to selection effects.   \citet{Giom15} postulate that blazars missing redshifts should be mostly high-luminosity HSP objects, in contrast to the fact that the low-energy SED peak correlates with luminosity. Therefore, measuring the redshift of a sizeable fraction of them could test this hypothesis by allowing   the luminosity to be measured.

  BL Lac redshifts can be estimated under the assumption that the host galaxy is a standard candle. The studies of BL Lac host galaxies have shown that BL Lacs are hosted in giant elliptical galaxies with absolute magnitude distribution well fitted by a Gaussian peaked at M$_{\rm R}$ $\sim$ $-$22.8 with FWHM of 1 magnitude \citep[see e.g.][and references therein]{Sbar05}. Therefore the redshift can be estimated either from optical images 
 \citep[see e.g.][]{Falo96,FaKo99,Nil03} or from optical spectra \citep{Sbar06}. The non-detection of the host galaxy or of its absorption features allows a lower limit to be set on the redshift of the source \citep[see review by][for further discussion and references]{Falo14}. A photometric method to derive limits on the redshifts of  BL Lacs   \citep{rau12} is based on the absorption of UV photons from a BL~Lac object by  the neutral hydrogen along our line of sight causing a clear attenuation in the flux at the Lyman limit (912 \AA). This dropout can be successfully used to measure the redshift of the BL Lac object if it is located at high redshift ($\ge 1.3$). More frequently for spectroscopy, if no intrinsic spectral feature is detected, a firm lower limit on the redshift may be set by the detection of an absorption system (usually MgII  doublets with wavelengths $\lambda_1$=2796.3 \AA~and $\lambda_2$=2803.5 \AA) along the line of sight towards the source.  It should also be noted that in certain cases conflicting redshift values are reported in the literature even for objects for which high S/N spectra have been obtained. For example, the redshift of 1ES 0502+675 has been reported as $z$ = 0.416 \citep{Landt02} and as $z$ = 0.314 \citep{Sca99}. Similarly, the redshift of PMNJ0816-1311 has been reported as $z$ = 0.046 \citep{Jon04,Jon09} and as $z$ > 0.288 \citep{Pit14}. Details on these and other cases are reported in Appendix A.

  Gamma-ray blazars are the main extragalactic targets for CTA and high-confidence spectroscopic redshifts are needed for them. The planning of CTA observations by the CTA Consortium is currently under way. It is therefore of great importance to start acquiring highly reliable redshifts for a large fraction of the AGN sources detected with {\em Fermi}-LAT that are likely to be detected with CTA. Such a redshift measurement campaign is recognised as necessary support for the CTA Key Science Programme (KSP) on AGN \citep{Sctashort19}. We thus initiated this redshift-measuring campaign by carrying out observations at different facilities to which we have access, and we report the first results here.
  
  This paper is organised in the following way:  the sample selection is presented in Section 2; the observing strategy in Section 3;  the observations, data reduction, and analysis in Sections 4 to 6; and the   discussion and conclusions in Section 7. For all calculations, we used a cosmology with $\Omega_{\rm M}$=0.27, $\Omega_{\rm \Lambda}$= 0.73, and H$_0$ = 70 km s$^{-1}$ Mpc$^{-1}$. All wavelengths are in air.  All magnitudes are in the AB system.

\section{Sample selection}

 The CTA will detect several hundreds of blazars in the VHE band \citep{Sctashort19}. It is expected that many of these blazars also emit  gamma rays at lower energies, in the energy range currently covered by the {\em Fermi}-LAT. Therefore, it is possible to  use the {\em Fermi}-LAT catalogues to identify a population of blazar candidates for CTA. The 3FHL  catalogue \citep{Fer3FHL17} is particularly interesting because it contains the spectral information, averaged over 7 years of its all-sky survey, for the harder and brighter sources detected by {\em Fermi}-LAT. The catalogue considers only photons above 10 GeV, which is very close to the energy threshold of CTA, and contains 1556 sources. The vast majority of them (1212) are blazars. The 3FHL blazars comprise 172 FSRQs, 750 BL Lacs, and 290 BCUs. 
A redshift value, if available, is provided for each source, but the information on its origin is not given. Conversely spectroscopic lower limits are not part of the catalogue. Among blazars in the 3FHL catalogue 95\% of FSRQs have a known redshift. Conversely 46\% of the BL Lacs and only 10\% of BCUs have a known redshift. We therefore focused on the 1040 BL Lacs and BCUs, of which only 373 (36\%) have a redshift in the 3FHL catalogue. 

We performed Monte Carlo simulations using the Gammapy\footnote{\url{https://gammapy.org}} software \citep{gammapy:2017,gammapy:2019} to estimate the minimal observation time necessary to detect at 5$\sigma$ each of these 1040 3FHL BL Lacs and BCUs with the North or South CTA array, depending on the declination of the source. We used publicly available CTA performance files\footnote{\url{https://www.cta-observatory.org/wp-content/uploads/2019/04/CTA-Performance-prod3b-v2-FITS.tar.gz}}. For each source, the average energy spectrum reported in 3FHL was extrapolated to very high energies and an intrinsic exponential cutoff at 3 TeV in the comoving frame was assumed
 in order to simulate the spectral curvature expected at these energies \citep{CTAcons20}\footnote {However,  in most cases the exact value of this cutoff has only a marginal effect on the estimation of the observation time required to reach a 5$\sigma$ detection as the detection significance is dominated by low-energy events.}.
 To take into account the energy and redshift-dependent absorption of gamma rays due to their interaction with the EBL, the spectral model was multiplied by exp(-$\tau$(E,z)), where $\tau$(E,z) is the gamma-gamma optical depth provided by \citet{Dom11}, E is the gamma-ray energy, and $z$ the source redshift from the 3FHL catalogue. For sources without a reported redshift in 3FHL, a value of $z$ = 0.3,  similar to $z_{\rm med}$ = 0.33 \citep{Shaw13} and to $z_{\rm med}$ = 0.285 \citep{Pena20} for BL Lacs, was considered.

 Sources were selected in a two-step process. In the first step a literature review or the analysis of publicly available archived spectra was performed for the 221 sources expected to be detected with CTA in less than 50 hours from the simulations defined above. This condition allowed us to reduce the number of sources for which the literature review was needed. During this check we examined published results and publicly available spectra for the selected sources. Among the sources having a reported redshift in 3FHL, 13 incorrect or unreliable redshift values were identified. These values were discarded because either we could not identify the features in the publicly available spectra or because they were contradicted or not confirmed by later spectra with much higher S/N (see Appendix A for details). In 2 out of the 13 cases, we associated instead published spectroscopic lower limits.
 Following the same procedure, among sources with no redshift reported in the 3FHL catalogue, a reliable spectroscopic redshift was assigned for 7 sources and a spectroscopic lower limit was associated with 12 sources. This process resulted in the revision of 32 redshift values (see the list of  sources in Appendix A). We note that the new lower limit values we obtained were adopted as redshifts whenever they were greater than our chosen value $z$ = 0.3; conversely, for smaller values we used $z$ = 0.3.
In the second step, using the revised redshift values, simulations were reprocessed. We then selected the sample of 165 sources without redshift measurement that are expected to be detectable in less than 30 hours if they are in the average spectral state reported by 3FHL, and in a significantly lower time if they are in a flaring state.
 
 As an early effort to determine the redshift of the sources in this sample, we extracted the 19 sources that we observed for this paper. The criteria used for this selection are explained in the next section.

\section{Observing strategy}

 We discuss here the strategy employed in deriving our observing campaign, the results of which are reported for the first time in this paper. 
   The goal is to obtain spectroscopic redshifts or lower limits for the highest possible number of sources in the sample. We plan to release our results as they become available so that they can be used to update the CTA Consortium observing programme \citep{Sctashort19}. These observations will also serve the  astronomical community at large as having confirmed redshifts for a larger sample of sources will help scientists who are investigating the properties of blazars and of their emission.
  
 To pursue this goal we have devised an observing programme aimed at constraining the redshift of these sources through deep imaging and spectroscopic observations. The goal of the imaging observations is to search for the extension of the source profile due to the host galaxy, and the first results will be reported in a follow-up paper (Fallah Ramazani et al. in prep.). The goal of the spectroscopic observations is to search for stellar absorption features of the host galaxy that are usually overwhelmed by the non-thermal continuum of the jet. As the host galaxies are usually luminous ellipticals \citep{Urr00}, the main features that we expect are the CaHK doublet, Mgb and NaID. Emission lines (especially [OII], [OIII], H$\alpha$, and N[II]) are only rarely detected. In all cases, EWs of about 5 \AA~or less are expected.
 To reach this goal we require that each spectrum has a spectral resolution $\lambda / \Delta \lambda$ of at least of a few hundred (if possible $\sim$1000), and an average  S/N  of $\sim$ 100 per pixel. The combination of these two constraints is extremely powerful. On the one hand, the imaging detection of the host galaxy is a clear indication of the likelihood of obtaining a redshift measurement. It has been shown for a sample of 100 X-ray detected BL Lacs that to date 90\% of the 62 targets with a detected host galaxy \citep{Nil03} have  a spectroscopic redshift from spectroscopic programmes \citep[see e.g.][]{Pena20,Pai20}, while more than 80\% of the unresolved sources still do not have redshift values. On the other hand, spectra at S/N  $\sim$100 and resolution $\sim$ 1000 allow  the detection of  weak host-galaxy features with EWs smaller than 5 \AA~and of intervening absorption systems \citep[see e.g.] [] {Pit14}. If the instrument we are using cannot provide us with spectra having both of these properties, we choose configurations that allow us to obtain at least one of them.

 We also performed a comprehensive literature search on our targets looking for previous spectroscopic results and for evidence of extension in archival and published data \citep[e.g. the Two Micron All sky Survey (2MASS) Extended Source Catalogue;][]{Skr06}. The results were classified in terms of reliability based on available information (images, plots, data). A source with low S/N spectroscopy and a tentative redshift value is a high-priority target.
 This allows us to concentrate our early efforts on promising and relatively uninvestigated sources. Conversely, if we find that a source  already has at least one deep and featureless spectrum and/or is not extended, it is classified as a low-priority target. A possible option for these sources is to trigger a spectroscopic observation during an epoch of low optical activity in order to take advantage of the improved S/N due to the lower non-thermal foreground.

\begin{sidewaystable*}
\small
  \caption{\label{tabobs1}List of observed sources and parameters of the observations. All sources were observed once with the exception of 1RXS J015658.6$-$530208, which was observed twice. The NTT/EFOSC2 spectra were all taken with Gr 6, except for NVSSJ151148$-$051345 (Gr 14) and NVSSJ152048$-$034850 (Gr 8). The sources with a $\dagger$ symbol are listed in the BZCAT catalogue \citep{bzcat15}.}
\centering
\begin{tabular}{lccccclcclll}
\hline\hline
3FHL name &   4FGL Name & Source name  & Ext. & RA & Dec  &   Telescope/ &   Slit            & Start Time & Exp.  & Airm. & Seeing      \\
         &   &    &      &     &      &   Instrument &  (\arcsec)  &  UTC       &  (sec)   &         & (\arcsec)      \\  
 (1) & (2)& (3) & (4) & (5) & (6) & (7) & (8) & (9)  &(10) & (11) & (12) \\ 
\hline
3FHL J0114.9$-$3359 & 4FGL J0114.9$-$3400  &1RXS J011501.3$-$340008$^{\dagger}$      & N & 01 15 01.6 & -34 00 27 & NTT/EFOSC2 & 1.5 & 2019-06-23 08:45:38 & 5400 & 1.23 & 0.8 \\
 3FHL J0156.7$-$5302   & 4FGL J0156.9$-$5301  & 1RXS J015658.6$-$530208$^{\dagger}$      & N & 01 56 58.0 & -53 01 60 & SALT/RSS   & 2.0 & 2019-11-24 23:01:16 & 2250 & 1.33 & 1.2 \\
3FHL J0156.7$-$5302 & 4FGL J0156.9$-$5301  & 1RXS J015658.6$-$530208$^{\dagger}$     & N & 01 56 58.0 & -53 01 60 & SALT/RSS   & 2.0 & 2019-11-26 22:34:09 & 2250 & 1.27 & 1.4 \\
 3FHL J0209.3$-$5229  & 4FGL J0209.3$-$5228 & 1RXS J020922.2$-$522920$^{\dagger}$      & N & 02 09 21.6 & -52 29 23 & SALT/RSS   & 2.0 & 2019-12-23 21:15:36 & 2220 & 1.28 & 1.4 \\
3FHL J1443.9$-$3908  & 4FGL J1443.9$-$3908  & PKS 1440$-$389$^{\dagger}$                       & Y & 14 43 57.2 & -39 08 40 & NTT/EFOSC2 & 1.5 & 2019-06-22 23:21:10 & 4500 & 1.10 & 1.1 \\
3FHL J1457.8$-$4642   & 4FGL J1457.8$-$4642  & PMN J1457$-$4642    & Y & 14 57 41.8 & -46 42 10 & NTT/EFOSC2 & 1.5 & 2019-06-23 23:15:26 & 2700 & 1.18 & 1.6 \\
3FHL J1511.8$-$0513   & 4FGL J1511.8$-$0513  & NVSS J151148$-$051345        & N & 15 11 48.5 & -05 13 47 & NTT/EFOSC2 & 1.5 & 2019-06-23 02:03:48 & 2700 & 1.10 & 0.9 \\
3FHL J1520.7$-$0348  &  4FGL J1520.8$-$0348  & NVSS J152048$-$034850$^{\dagger}$          & N & 15 20 48.9 & -03 48 51 & NTT/EFOSC2 & 1.5 & 2019-06-23 03:00:00 & 2700 & 1.15 & 1.0  \\
3FHL J1532.7$-$1319  & 4FGL J1532.7$-$1319  & TXS 1515$-$273                          & Y & 15 17 59.5 & -27 32 51 & NTT/EFOSC2 & 1.5 & 2019-06-23 00:51:57 & 3600 & 1.03 & 0.7  \\
3FHL J1539.7$-$1127  & 4FGL J1539.7$-$1127 & PMN J1539$-$1128                    & N & 15 39 41.2 & -11 28 35 & NTT/EFOSC2 & 1.5 & 2019-06-24 00:19:35 & 5400 & 1.14 & 1.6 \\
 3FHL J1548.4$+$1456  & 4FGL J1548.3$+$1456  & WISE J154824.39$+$145702.8  & Y & 15 48 24.4 & +14 57 03 & Keck/ESI   & 1.0 & 2018-05-13 10:34:13 & 3600 & 1.01 & 0.6  \\
3FHL J1637.8$-$3448 & 4FGL J1637.8$-$3449 & NVSS J163750$-$344915    & N & 16 37 51.0 & -34 49 15 & NTT/EFOSC2 & 1.5 & 2019-06-24 02:01:38 & 2700 & 1.04 & 0.7 \\
3FHL J1838.8$+$4802 & 4FGL J1838.8$+$4802 & GB6 J1838$+$4802$^{\dagger}$ & Y & 18 38 49.2 & +48 02 34 & Keck/ESI   & 1.0 & 2018-05-13 12:28:22 & 3900 & 1.15 & 0.6 \\
3FHL J1841.3$+$2909 &  4FGL J1841.3$+$2909  & MITG J184126$+$2910  & Y & 18 41 21.7 & +29 09 41 & Keck/ESI   & 1.0 & 2018-05-13 11:41:46 & 2400 & 1.11 & 0.5 \\
3FHL J1842.4$-$5841  & 4FGL J1842.4$-$5840  & 1RXS J184230.6$-$584202   & N & 18 42 29.8 & -58 41 56 & NTT/EFOSC2 & 1.5 & 2019-06-23 04:12:40 & 4500 & 1.17 & 1.0 \\
3FHL J1958.3$-$3011 & 4FGL J1958.3$-$3010 & 1RXS J195815.6$-$301119$^{\dagger}$         & Y & 19 58 14.9 & -30 11 11 & NTT/EFOSC2 & 1.5 & 2019-06-23 05:38:55 & 1800 & 1.03 & 1.0 \\
 3FHL J2001.2$+$4353 & 4FGL J2001.2$+$4353  & MAGIC J2001$+$435$^{\dagger}$ & Y & 20 01 12.9 & +43 52 53 & Keck/ESI   & 1.0 & 2018-05-13 13:38:54 & 3480 & 1.13 & 0.6 \\         
3FHL J2036.9$-$3328 & 4FGL J2036.9$-$3329  & 1RXS J203650.9$-$332817        & N & 20 36 49.5 & -33 28 30 & NTT/EFOSC2 & 1.5 & 2019-06-25 08:07:13 & 3600 & 1.04 & 0.8 \\
3FHL J2131.0$-$2746  & 4FGL J2131.0$-$2746  & RBS 1751$^{\dagger}$  & N & 21 31 03.3 & -27 46 58 & NTT/EFOSC2 & 1.5 & 2019-06-25 09:26:18 & 3600 & 1.07 & 1.5 \\ 
3FHL J2324.7$-$4040 &  4FGL J2324.7$-$4041 & 1ES 2322$-$409$^{\dagger}$  & N & 23 24 44.7 & -40 40 49 & NTT/EFOSC2 & 1.5 & 2019-06-23 06:18:39 & 7200 & 1.31 & 0.8 \\
\hline
\hline

\end{tabular}
\tablefoot{The columns are (1) 3FHL Name, (2) 4FGL Name, (3) Source Name, (4) Extension Flag, (5) Right Ascension (J2000), (6) Declination (J2000), (7) Telescope and Instrument, (8) Slit Width in arcsec, (9) Start Time of the observations, (10) Exposure Time, (11) Average Airmass, and (12) Average Seeing.}
\end{sidewaystable*}

\section{Observations and data reduction}


 Data were collected on 19 blazars using three different instruments at three
facilities for a total observation time of about 17.5 hours between May 2018 and
November 2019. Observations were performed using the Echellette Spectrograph and Imager \citep[ESI;][]{Shei12} installed on the Keck II telescope at the Keck observatory, with the Robert Stobie Spectrograph \citep[RSS;][]{Burgh03} on the Southern African Large Telescope (SALT) at the South African Astronomical Observatory and with the ESO Faint Object Spectrograph and Camera \citep[EFOSC2;][]{Buz84} on the New Technology Telescope at La Silla Observatory. The Keck II  and SALT telescopes  respectively have 10-meter and 11-meter diameter primary mirrors, while the NTT primary mirror  is substantially smaller (3.5 meters in diameter).
The list of observed sources together with the details of the observations are given in Table \ref{tabobs1}.

\subsection{Keck/ESI}

 The ESI spectrograph is a visible-wavelength faint-object imager and single-slit spectrograph; it has been in operation at the Cassegrain focus of the Keck II telescope since 1999. We used it in its main spectroscopy mode, the echellette mode, which has a single-shot wavelength coverage of 3900 - 10000 \AA, a throughput up to 28 \%, and spectral resolution $\lambda / \Delta \lambda \sim$ 10000. Two short observations of the flux standard HD165459 with slit widths of 1 and 6 arcsec were performed at the end of the night to allow for flux calibrations.

The data reduction was performed using the XIDL pipeline\footnote{\url{https://github.com/profxj/xidl}} based on the Interactive Data Language (IDL\footnote{\url{http://www.harrisgeospatial.com}}) software . We used the pipeline to perform bias subtraction, flat-field division, wavelength calibrations, cosmic ray subtraction and spectral extraction. We then performed order merging and flux calibration using our own procedures under IDL.
Telluric corrections were performed using molecfit \citep{Sme15, Kau15}. The spectra were dereddened using the maps of \citet{Sch11} and the extinction curve of \citet{Fitz99}.

\subsection{SALT/RSS}

 The RSS is   SALT's main   instrument; it is a complex multimode instrument with a wide range of capabilities. We used it in Long Slit Spectroscopy (LSS) mode with the PG0900 grating and a 2 arcsec slit. This configuration is sensitive between 4500 and 7500 \AA~and it has a throughput greater than 20 \% \citep{Kob03}.

  We reduced the spectra using PySALT \citep{Cra10} accounting for cross-talk, bias, gain, and flat-field correction. Wavelength calibration was performed using standard IRAF routines \citep{Tod86},  while cosmic ray cleaning and flux calibration were performed under IDL. To estimate the spectral resolution we extracted the sky spectrum and fitted ten isolated sky lines across the spectrum with Gaussian functions. This analysis produced an approximate value of the spectral resolution $\lambda / \Delta \lambda$ of about 1000. To perform flux calibration, we used observations of the standard star HILT600 taken on November 26, 2019, with a 4 arcsec slit. These data were reduced in the same way as the data of the target. Other standard star observations taken near the end of December 2019 were of  much lower quality and were therefore discarded.  If more than one high-quality observation was obtained, an average spectrum was produced. Telluric and reddening corrections  were performed as described in section 4.1.

 SALT has a moving, field-dependent, and under-filled entrance pupil, which makes absolute flux calibration difficult to achieve to a good degree of accuracy \citep[see e.g.] [] {Buc18}. We therefore obtained near-contemporary photometric observations for both targets. 
 
 Our first target, 1RXS J015658.6-530208, was observed using  the Ultraviolet/Optical Telescope \citep[UVOT;][170--600 nm]{Rom05} on board the {\em Neil Gehrels Swift Observatory} \citep{Geh04} on November 30, 2019.  The UVOT instrument observed in the optical ($u$, $b$, and $v$) photometric bands \citep{Poo08,Bre10} with exposures of 305 s, 300 s, and 305 s for the $u$, $b$, and $v$ filter, respectively. We analysed the data using the \texttt{uvotsource} task included in the \texttt{HEAsoft} package (v6.22). Source counts were extracted from a circular region of 5 arcsec radius centred on the source, while background counts were derived from a circular region of 20 arcsec radius in a nearby source-free region. The results are presented in Table \ref{tabUVOT}.
 
 The second target, 1RXS J020922.2-522920, was observed  using the REM Optical Slitless Spectrograph (ROSS2) at the REM telescope \citep{Zer01, Cov04}, a robotic telescope located at the ESO Cerro La Silla observatory (Chile).  With the ROSS2 instrument we obtained two 240 s integration images of the target in $g$, $r$, and $i$ filters on five separate dates between the end of December 2019 and the beginning of January 2020. All raw optical frames obtained were reduced following standard procedures. Instrumental magnitudes were obtained via aperture photometry and absolute calibration was performed by means of secondary standard stars in the field reported by the American Association of Variable Star Observers Photometric All-Sky Survey (APASS) catalogue\footnote{\url{https://www.aavso.org/apass}}. The results are shown in Table \ref{tabREM2}.

\subsection{NTT/EFOSC2}

The EFOSC2 is a versatile multimode instrument that is particularly efficient in low-resolution spectroscopy. In order to obtain a wide wavelength coverage with good sensitivity and reasonable resolving power we selected  Grism 6 for 11 of our 13 targets. Grism 6 is sensitive in the range 3860-8070 \AA, but with  low spectroscopic resolution $\lambda / \Delta \lambda \sim$ 400. For the remaining two, NVSSJ151148-051345 and NVSSJ152048-034850, we selected Grism 14 (3095-5085 \AA; $\lambda / \Delta \lambda \sim$ 550) and Grism 8 (4320-6360 \AA ; $\lambda / \Delta \lambda \sim$ 660), respectively, which are more adapted to investigating previously reported detections of MgII absorbers in their spectra. The throughputs of the grisms\footnote{\url{https://www.eso.org/sci/facilities/lasilla/instruments/efosc/images-/GrismAllEfficiency2004.jpg}} are between 20 \% and 30 \% . In all cases we used a 1.5 arcsec slit; the same slit was also used  for standard stars.

The observations were performed during three nights from June 22 to June 25, 2019 (see Table \ref{tabobs1}), with variable atmospheric conditions. Data reduction was performed using the ESO/EFOSC2 pipeline version 2.3.3 and esorex version 3.13.2. The pipeline performs  bias subtraction, flat-field correction, and wavelength calibration using daytime calibration files. Cosmic ray subtraction was performed under IDL. Subsequently flux calibration was performed using the standard stars observed during the run. These steps were performed for each independent frame; we then averaged the extracted spectra to obtain the final flux calibrated spectra. During flux calibration we discovered distortions in the spectral shape of the sources, likely due to the presence of clouds. We corrected them using spectra of stars that were put in the slit of some of our targets. Telluric and reddening corrections were performed as described in section 4.1.

\begin{table*}
\caption{\label{tabres1} Analysis results for all the observed sources. As the redshift of 1RXS J184230.6$-$584202 is based on a low-confidence detection of the CaHK feature, for this source we also present  the results of a simple power-law fit. The spectral bin width is 4 \AA~for the sources observed with EFOSC2; 1 \AA~for MITG J184126$+$2910, MAGIC J2001$+$435,  1RXS J015658.6$-$530208, and 1RXS J020922.2$-$522920; and 2 \AA~for WISE J154824.39$+$145702.8.}
\centering
\begin{tabular}{lccccccc}
\hline\hline

 Source name  & S/N &  R$_{\rm c}$(BL Lac) & Redshift   & Flux Ratio  & R$_{\rm c}$(gal) & M$_{\rm R}$ & Slope   \\
              &     &   (obs)              &            &            &  (fit)            &   (gal)     &         \\  
   (1)  & (2) & (3)    &  (4)  &  (5)   &  (6)      &  (7)   &  (8) \\      
\hline
1RXS J011501.3$-$340008   & 20   & 19.4$\pm$0.1    &  0.4824  $\pm$  0.0007    &  0.9$\pm$0.2   & 19.7$\pm$0.3   &  -23.3     & -1.8$\pm$0.4  \\ 
1RXS J015658.6$-$530208  &  100 &  17.0$\pm$0.3   &  0.3043 $\pm$  0.0004      & 3.2$\pm$1.3    &  18.3$\pm$0.5 &  -22.7    & -1.4$\pm$0.2  \\
1RXS J020922.2$-$522920     & 160 &  15.4$\pm$0.2   &  0.2110  $\pm$  0.0002        &  6.0$\pm$1.8   &  17.1$\pm$0.4 &  -23.2      & -1.5$\pm$0.2  \\
PKS 1440$-$389    & 230  & 15.0$\pm$0.1     &  0.1385 $\pm$  0.0005 &  8.6$\pm$0.7   & 16.8$\pm$0.2  &  -22.4      & -1.5$\pm$0.2   \\ 
PMN J1457$-$4642     &  45   & 17.5$\pm$0.2  &  0.1116 $\pm$  0.0002  & 0.3$\pm$0.2    & 16.7$\pm$0.1  &  -22.2  & -2.3$\pm$0.2   \\      
NVSS J151148$-$051345   &   23   & 17.5$\pm$0.1$^{\dagger}$  &  $\ge$ 0.4480  $\pm$  0.0003& ---     &   ---       &   ---       &  -2.5$\pm$0.2   \\
NVSS J152048$-$034850   &   43   &    17.2$\pm$0.1     & $\ge$ 0.8680 $\pm$  0.0002 & ---     &  ---      &    ---     &  -1.6$\pm$0.1   \\
TXS 1515$-$273   & 160   & 15.6$\pm$0.1 &  0.1284 $\pm$  0.0003 &  3.0$\pm$0.3   & 16.6$\pm$0.2   &  -22.4   & -1.6$\pm$0.1    \\
PMN J1539$-$1128 &   80     &  17.8$\pm$0.1  & ---      &  ---        &   ---             &  ---   &  -2.1$\pm$0.3 \\
WISE J154824.39$+$145702.8 & 42 & 18.2$\pm$0.1    &  0.2308 $\pm$  0.0002   & $\le 0.03$ & 18.2$\pm$0.1   &  -22.3    & ---   \\   
NVSS J163750$-$344915   &   80   & 17.5$\pm$0.1   & ---    &   ---    &   ---   &  ---   & -2.3$\pm$0.2 \\
GB6 J1838$+$4802 &   250  & 15.8$\pm$0.1 &  ---   &  ---    &  ---   & ---   & -1.1$\pm$0.1 \\
MITG J184126$+$2910   & 100   & 17.4$\pm$0.1 &  0.2883  $\pm$  0.0003  & 2.6$\pm$0.5   & 18.2$\pm$0.3    & -22.9  & -1.5$\pm$0.1   \\
1RXS J184230.6$-$584202   &  35    & 18.3$\pm$0.2   &  0.421$^*$  &  1.7$\pm$0.3   & 19.5$\pm$0.2    & -23.0 & -2.4$\pm$0.3    \\
1RXS J184230.6$-$584202  &  35    & 18.3$\pm$0.2  &  ---   &    ---   &     ---   &  ---  &  -1.6$\pm$0.3   \\
1RXS J195815.6$-$301119  & 110   & 15.8$\pm$0.1   &  0.1190 $\pm$  0.0003  & 1.2$\pm$0.2 & 16.7$\pm$0.2   &  -22.1   & -1.9$\pm$0.2    \\
MAGIC J2001$+$435    & 105    & 15.9$\pm$0.1  & 0.1739 $\pm$  0.0004 & 5.0$\pm$0.4    & 17.0$\pm$0.2 &  -22.4     & -1.0$\pm$0.1   \\ 
1RXS J203650.9$-$332817 &   19    &  17.8$\pm$0.1   &  ---     &  ---     &  ---  &  ----   & -1.6$\pm$0.2 \\
RBS 1751 &   40 & 16.9$\pm$0.1   & $\ge$0.618$^*$ &  --- &  ---  &  ---  & -1.6$\pm$0.2    \\
1ES 2322$-$409   & 210   & 16.0$\pm$0.2  & 0.1736  $\pm$  0.0008 &  9.5$\pm$0.6   & 17.7$\pm$0.2 &  -22.1   &  -1.6$\pm$ 0.4   \\

\hline
\hline
\end{tabular}
\tablefoot{The columns are (1) Source Name; (2) Median signal-to-noise ratio per spectral bin measured in continuum regions; (3) R$_{\rm c}$, Cousins magnitude of the BL Lac spectrum corrected for reddening, telluric absorption, and slit losses with errors. Slit losses were estimated using an effective radius r$_e$=10 kpc for all sources except for MAGIC J2001$+$435 for which we used r$_e$ = 6.8 kpc \citep{Ale14b}; (4) Redshift or lower limit with error, (5) Flux Ratio jet/galaxy at 5500 \AA~in rest frame; (6) R$_{\rm c}$, Cousins magnitude of the galaxy with the same corrections as in column (3); (7) Absolute R Magnitude of the galaxy, the errors are the same as those in  column (6); (8) Power-Law Slope with errors.  If the entry is unknown, the legend is `---'.

$^{\dagger}$ U Magnitude

$^*$ Uncertain Redshift

}
\end{table*}

 \begin{table*}
\caption{\label{tabeqw} Equivalent widths in \AA~of the absorption features detected in the spectra at the measured redshift for each source. The CaFe feature of 1RXS J195815.6$-$301119 is likely contaminated by Galactic NaID.}

\centering
\begin{tabular}{lccccc}
\hline\hline

Source Name &  CaHK & CaIG & Mgb & CaFe & NaID  \\
            &       &      &     &       &     \\
 (1)  &  (2)   &  (3) & (4) & (5) &  (6)  \\        

\hline

1RXS J011501.3$-$340008         & 6.4$\pm$1.6   & 2.5$\pm$0.7 &   ---                 & ---                       &  --- \\
1RXS J015658.6$-$530208        & 2.3$\pm$0.3   & 0.8$\pm$0.1 & 0.9$\pm$0.2   & ---                        &  ---   \\
1RXS J020922.2$-$522920        & 1.9$\pm$0.3   & 0.5$\pm$0.1 &  0.5$\pm$0.2  & ---                        &  0.5$\pm$0.1 \\
PKS 1440$-$389                         & 0.6$\pm$0.1   & 0.5$\pm$0.1 &  ---                   & ---                        & 0.4$\pm$0.1 \\
PMN J1457$-$4642                    & 11.8$\pm$0.8  & 4.8$\pm$0.7 & 11.0$\pm$1.2  & ---                        & 6.8$\pm$0.9 \\
TXS 1515$-$273                         & 1.2$\pm$0.2   & 0.3$\pm$0.1 &  0.3$\pm$0.1   & ---                        & 0.7$\pm$ 0.2 \\
WISE J154824.39$+$145702.8  & 25.0$\pm$0.9 & 7.2$\pm$0.3 &  6.5$\pm$0.3    & 2.5$\pm$0.3       & 4.8$\pm$0.2 \\
MITG J184126$+$2910              & 1.6$\pm$0.3   & 0.8$\pm$0.2 &  0.7$\pm$0.2    &  0.6$\pm$ 0.2      &     --- \\
1RXS J184230.6$-$584202       & 3.6$\pm$1.0$^*$   &  ---                 &   ---                   &   ---                       &   --- \\
1RXS J195815.6$-$301119       & 3.3$\pm$0.3    & 1.2$\pm$0.3 &  5.3$\pm$0.6    & 2.4$\pm$0.2$^*$  & 5.5$\pm$0.4 \\
MAGIC J2001$+$435                & 2.2$\pm$0.2    &  ---                 &  1.0$\pm$0.2     & 0.8$\pm$0.1       &  0.4$\pm$0.2 \\
1ES 2322$-$409                        & 0.6$\pm$0.1   & 1.2$\pm$0.2 &  2.3$\pm$0.3      &  ---                       & 1.2$\pm$0.3 \\
\hline
\hline
\end{tabular}
\tablefoot{The columns are (1) Source Name, (2) Equivalent Width of the CaHK feature with errors, (3) Equivalent Width of the CaIG feature with errors, (4) Equivalent Width of the Mgb feature with errors, (5) Equivalent Width of the CaFe feature with errors, (6) Equivalent Width of the NaID feature with errors. If the feature is not detected, the legend is `---'. The detection of CaHK in 1RXS J184230.6$-$584202 is uncertain and it is flagged with an asterisk. }
\end{table*}

 \begin{table}
\small
\caption{\label{tabeqw2} Equivalent width in \AA~of the main emission features detected in the spectra at the measured redshift. The EWs of the emission features of WISE J1548.24.39+145702.8 are in Table \ref{WISEtab}}

\centering
\begin{tabular}{lccc}
\hline\hline
Source Name &      [OII]     &      [OIII]            & H$\alpha$ -  [NII] \\
            &                &      $\lambda$ 5007    &                     \\
  (1)  &   (2)     &   (3)           &   (4)     \\      
\hline
PMN J1457$-$4642               &  ---             & ---           &  -5.0$\pm$0.9  \\
TXS 1515$-$273                    & -0.8$\pm$0.1   & -0.8$\pm$0.1 &      NA           \\
1RXS J195815.6$-$301119   & ---            & ---            & -4.8$\pm$1.0 \\
MAGIC J2001$+$435            & ---            &  ---           & -0.4$\pm$0.1 \\
\hline
\hline
\end{tabular}
\tablefoot{The columns are (1) Source Name, (2) Equivalent Width of the [OII] feature with errors, (3) Equivalent Width of the [OIII]$\lambda$ 5007 feature with errors, (4) Equivalent Width of the H$\alpha$-[NII] complex  with errors. If the feature is not detected, the legend is  `---'.}
\end{table}

 \section{Redshift measurement and estimation of the blazar total emission} 
  
    The optical spectrum of a blazar is a combination of non-thermal jet emission, AGN activity (thermal and non-thermal), and stellar emission of the host galaxy, usually an elliptical \citep{Urr00}. The jet emission has the form of a featureless power law  $f_{\lambda} \propto \lambda^{\alpha}$, which, as discussed above, is often much stronger than the host galaxy emission, making the host spectral features undetectable. Simulations \citep{Landt02,Pir07} have shown that when the  rest frame jet-to-galaxy ratio at 5500 \AA~is around 10 the features are already very difficult to detect.
  
   For each source we carefully searched for absorption or emission features that could be used to measure the redshift. When a possible feature was found we checked for the presence of other possible features at the same redshift. We then analysed the features in the following way. The spectra were normalised with cubic splines and the flux of each pixel  was integrated to determine the total EW of each line.  The uncertainties were estimated by taking the square root of the quadratic sum of the error spectrum and taking into account the errors of the continuum placement \citep[see the  Appendix in][]{Sem92}. The results are shown in Table \ref{tabeqw} and Table \ref{tabeqw2}. Only for WISE J154824.39$+$145702.8  did we measure the EWs of the rich emission line spectrum by Gaussian fitting (see Table \ref{WISEtab}).
  
 We considered two factors to estimate the uncertainty on redshift measurements: uncertainties in wavelength calibration and uncertainties in the position of the detected features. The dispersion  of the wavelength calibration in our spectra is always smaller than 0.5 \AA~from $\sim$ 4000 to $\sim$ 8000 \AA, which translates into a relative precision smaller than 6-12 $\times$ 10$^{-5}$ (18-36 km/s). Once the redshift was determined, we fitted Gaussian functions at the positions of the features found in each source listed in Tables 3 and 4, and took the variance of the fitted positions as the uncertainty. We then summed these with uncertainties in wavelength calibration and we obtained the total uncertainty estimates between 2 and 8 $\times$ 10$^{-4}$  listed in Table 2.
  
     After the redshift was determined, we modelled the spectrum with a combination of a power law describing the jet continuum and templates describing the elliptical galaxy emission \citep{Man01, Bru03}, adding Gaussian emission features when needed \citep{Pit14}.  For simplicity we used only one template per spectrum.
    The fit was performed in the rest frame using the MPFIT software \citep{Mar09} with two free parameters: the jet-to-galaxy ratio and the power-law slope. We estimated the goodness of fit from the value of the $\chi^2_{\rm dof}$. There are systematic differences between the flux calibrated spectra and the models we used. This is a common occurrence as, in general, polynomials are added to the spectral models in order to obtain an acceptable match with the calibrated flux \citep[e.g.][]{Cap17}.
    Given the weakness of our detected features in most cases we chose not to add polynomials as it would have led to overfitting. We thus added systematic errors to obtain error estimates. In some cases, the errors of the fit parameters were unphysically small, thus we independently fitted separate sections of the spectra and we estimated the errors from the differences between the resulting parameters. The results of these fits are presented in Table \ref{tabres1}.

   We also estimated the absolute magnitude of the detected host galaxies. To estimate the slit losses, we assumed the value of the effective radius of the host galaxy r$_e$ as 10 kpc for a de Vaucouleurs profile. The only detected host for which we had a photometric estimate of the effective radius is MAGIC J2001$+$435 for which we used r$_e$=6.8 kpc (from the measured r$_e$=2.4 arcsec). Within  the uncertainties, the magnitude we obtain for this host is compatible with the photometric value quoted in \citet{Ale14b}. Thus, we estimate that the uncertainties on the absolute magnitude are the same as the uncertainties on the measured magnitude. 
  The K-corrections were computed from the template spectra and we did not apply evolutionary corrections. The absolute magnitudes can be found in Table \ref{tabres1}.

   When the host galaxy was not detected, we fitted the spectrum with a power law with normalisation at the centre of the band and we estimated the errors fitting separate sections of the spectra as described above. The results are presented in Table \ref{tabres1}.

  \section{Sources and results}
 
 In the following we discuss the results of our observations for each of the sources.
 
\begin{figure*}
   \centering
 \includegraphics[width=6.8truecm,height=5.75truecm]{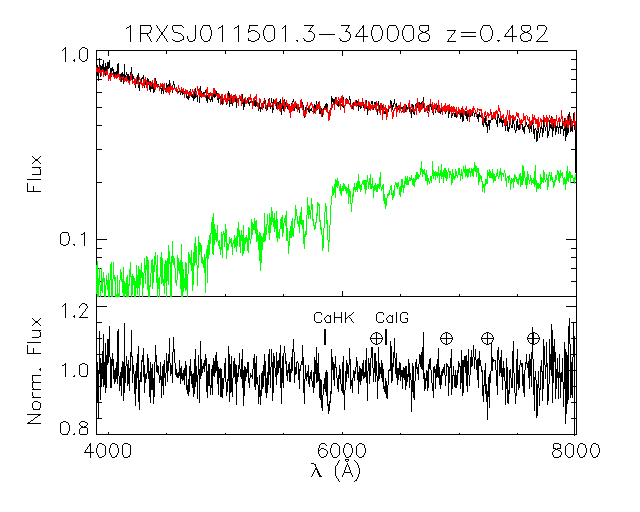}  \includegraphics[width=6.8truecm,height=5.75truecm]{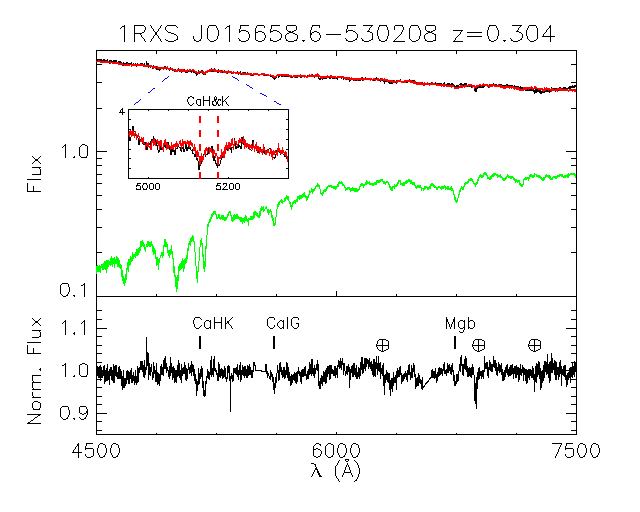} 
 \includegraphics[width=6.8truecm,height=5.75truecm]{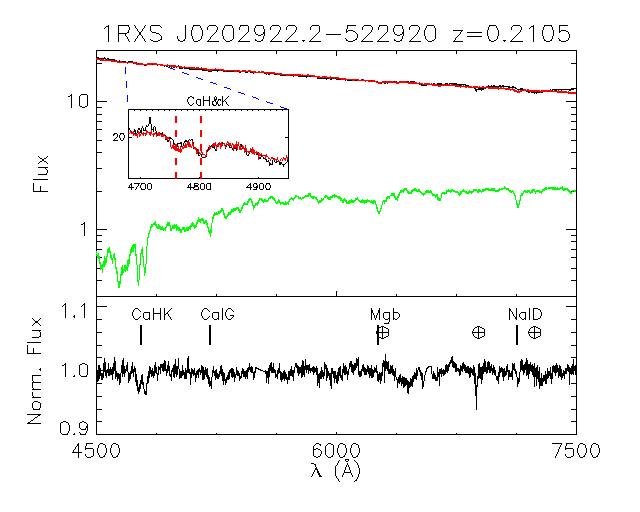}  \includegraphics[width=6.8truecm,height=5.75truecm]{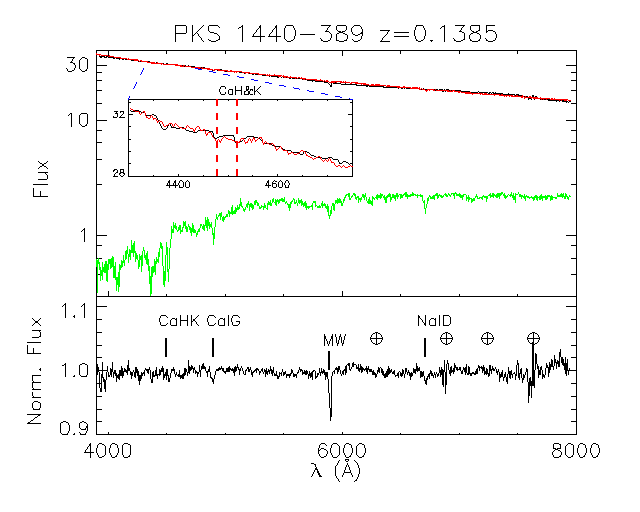}
  \includegraphics[width=6.8truecm,height=5.75truecm]{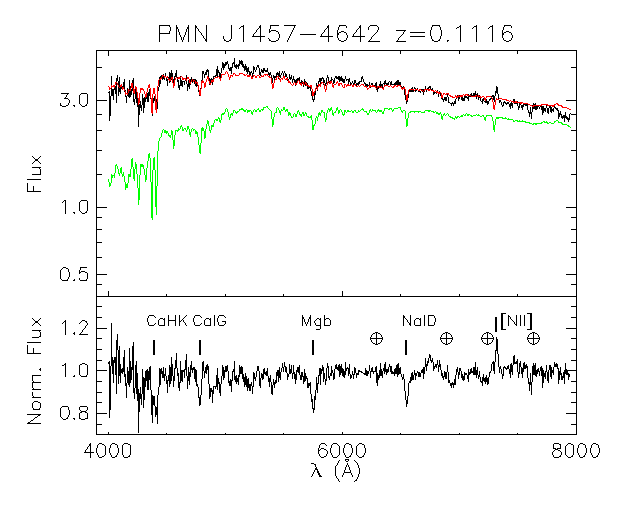}  \includegraphics[width=6.8truecm,height=5.75truecm]{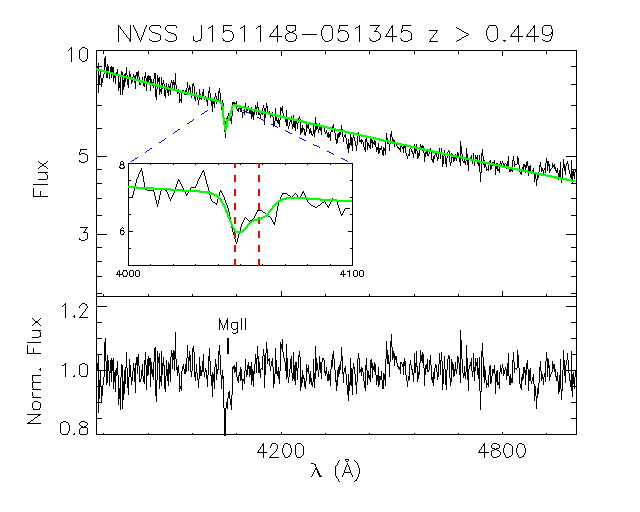}
 \includegraphics[width=6.8truecm,height=5.75truecm]{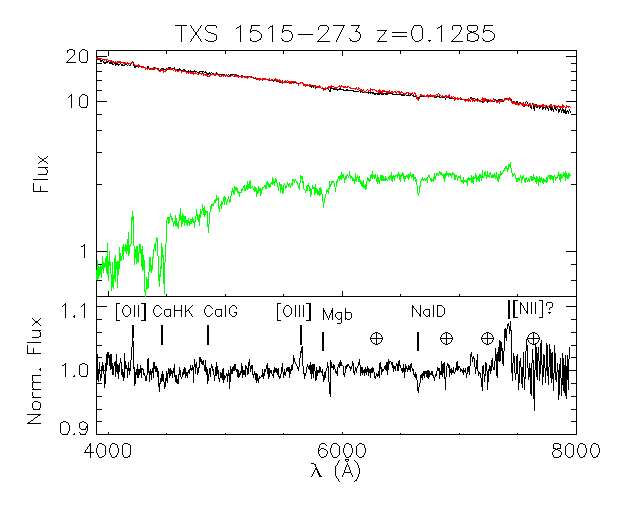}  \includegraphics[width=6.8truecm,height=5.75truecm]{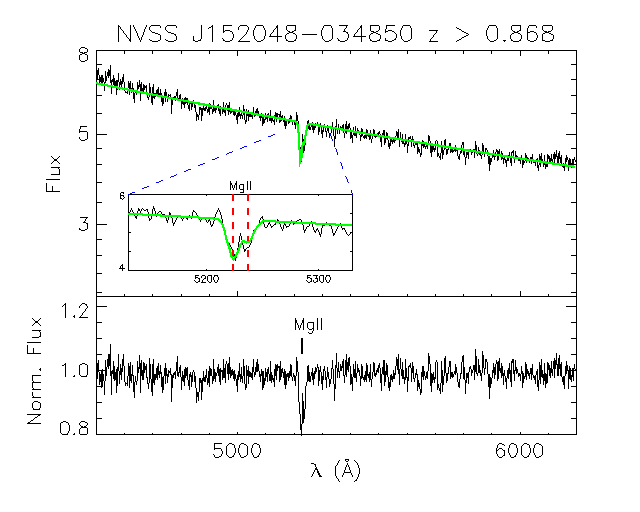}
   \caption{
   Flux-calibrated and normalized spectra of the first eight sources in Table \ref{tabobs1}. Each panel contains the spectrum, continuum, and galaxy model for a given source.  Each panel has  two parts. Upper:  Flux-calibrated and telluric-corrected spectrum (black) alongside the best fit model (red). The flux is in units of 10$^{-16}$ erg  cm$^{-2}$ s$^{-1} $\AA$^{-1}$. The elliptical galaxy component is shown in green. Lower:  Normalised spectrum with labels for the detected absorption features. Atmospheric telluric absorption features are indicated by the symbol $\oplus$ and Galactic absorption features are labelled  `MW'. }

\label{fig_spec1}
    \end{figure*}

\subsection{1RXS J011501.3$-$340008}

1RXS J011501.3$-$340008 was identified as an extreme HBL by \citet{Giom05}. While its redshift is unknown in the 3FHL, \citet{Pir07} report its redshift to be $z$ = 0.482 on the basis of a one-hour EFOSC2 spectrum taken at the ESO 3.6 m telescope.  The plot of the spectrum, shown in the Appendix of their paper, shows a medium- to low-quality spectrum. To assess the reliability of this result we downloaded the public data of the observation to reduce them. The resulting spectrum has a general power-law shape and displays a spectral break around $\lambda \sim $ 5900 \AA~that can be attributed to the CaHK feature at $z \sim $ 0.482. However its S/N is only $\sim$ 16 per 4 \AA~pixel. We tried to obtain a much longer integration with EFOSC2, but  we were able to obtain only an integration of 1 hour and 30 minutes (see Table \ref{tabobs1}).
The resulting spectrum has a median S/N = 20 and it has a general shape remarkably similar to the first   with a spectral break at the same wavelength. Given the similarity of the two spectra, we averaged them to measure the properties of the total emission (Fig. \ref{fig_spec1}, first row, left).
We detect the CaHK feature at 4$\sigma$ and the CaIG feature at 3.5$\sigma$ both at redshift $z$ = 0.4824 $\pm$ 0.0007. Given the stability of the source spectrum and despite the relatively low S/N we consider that this is a firm redshift. The spectral fit gives a bright host galaxy magnitude M$_{\rm{R}}$=-23.3 $\pm$ 0.2. Although a fit with a local template \citep{Man01} is satisfactory, the best results are obtained using a 2.5 Gyr Simple Stellar Population model \citep{Bru03}. This suggests the possibility of an anomalous star formation history for this object; further analysis on this subject is beyond the scope of this paper.

\subsection{1RXS J015658.6$-$530208}

  The BL Lac nature of 1RXS J015658.6$-$530208 was established by a low S/N featureless optical spectrum taken with the Goodman spectrograph at the Southern Astrophysical Research (SOAR) telescope  \citep{Lan15}. It was later classified as an EHBL \citep{Cos02,Fof19}.
 
  Our SALT/RSS observations were performed on November 24 and 26 2019. The transparency was good in both observations, and seeing was around  1.2 and 1.4 arcsec, respectively. The source was clearly detected in both observations at a median continuum S/N  of 100. Inspection of the average spectrum shows the presence of clear CaHK and CaIG features at $z \sim$0.304 (Fig. \ref{fig_spec1}, first row, right). This result is confirmed by the presence of a weaker Mgb feature at the same redshift, while the CaFe feature falls into residual telluric absorption and is therefore undetectable. The fit of the features gives a precise redshift value $z$ = 0.3043 $\pm$ 0.0004.
 
  We then compared the spectrum with the near-contemporaneous {\em Swift}/UVOT photometric points (see section 4.2.1). The {\em Swift}/UVOT fluxes are higher than the SALT/RSS values, while the slopes are comparable (see Fig. \ref{SALT_photo}).  We used the $v$ {\em Swift}/UVOT filter, whose bandwidth is completely contained in our spectrum, to rescale it. The ratio of the $v$ flux to the spectral flux in the corresponding range is 1.3 $\pm$ 0.2. We therefore multiplied the spectrum by 1.3 to match the UVOT photometry. 
A possible caveat to this analysis is the possibility of substantial variability (i.e. greater than our errors)  between the spectroscopic and photometric observation. To estimate this effect we fitted the weekly variability distribution of the Catalina survey \citep{Dra09} photometry of the source with a Gaussian function. The 2$\sigma$ width of the distribution is 0.24 magnitudes, which, summed in quadrature with the error of the flux ratio, pushes our errors to $\pm$ 0.3 which we use as the errors on the flux.
The fit with a power law plus  galaxy template model results in a bright host galaxy with $M_{\mathrm R}$=-22.7 $\pm$ 0.5.

 \subsection{1RXS J020922.2$-$522920}
 
 Spectra of 1RXS J020922.2$-$522920 have been published in the six-degree Field Galaxy Survey (6dF) \citep{Jon04,Jon09} and by \citet{Shaw13}.  From the first spectrum a redshift $z$ = 0.873 has been derived, while from the second spectrum a statistical lower limit $z  \ge$ 0.31 was proposed. We inspected both spectra and we could not detect any features in them; they also both have a low S/N.

 We observed it with SALT/RSS on 2019 December 28.  The source was clearly detected at a median continuum S/N  of 160.  Careful inspection of the spectrum revealed the presence of a clear CaHK feature at $z \sim$  0.211 (Fig. \ref{fig_spec1}, second row, left). This result is confirmed by the presence of weaker CaIG, Mgb, and NaID features at the same redshift, while the CaFe feature falls into residual telluric absorption and is therefore undetectable. Our spectroscopic redshift value is $z$ = 0.2110 $\pm$ 0.0002 and  is at odds with both of the previously published values; however, assuming the same spectral state, the features we report here could not have been detected in the previous spectra.

   We then compared the spectrum with the average ROSS photometric points (see section 4.2.1).
  The average ROSS fluxes are all consistent with the SALT/RSS spectrum within 0.1 magnitudes (see Fig. \ref{SALT_photo}). Concerning the effect of optical variability, we performed the same analysis as for 1RXS J015658.6-530208 using the Catalina survey light curve of the source. In this case the expected variability in one week (the time between the SALT/RSS observation and the average time of the ROSS observations) is 0.14 mag (2$\sigma$ level), again slightly increasing our errors to $\pm$ 0.2. We therefore kept   our original values. The host galaxy magnitude obtained from our fit is very bright at $M_{\mathrm R}$=-23.2 $\pm$ 0.4.

 \subsection{PKS 1440$-$389}

 PKS 1440$-$389, in addition to its strong GeV gamma-ray emission detected by {\em Fermi}-LAT, is also a bright TeV source \citep{Abda20}. A tentative redshift of $z$ = 0.069 has been reported from 6dF low S/N spectroscopy (Jones et al. 2004); we examined the spectrum, but could not find convincing spectral features at that redshift. Later, spectra with higher S/N (up to $\sim$ 80) taken by \citet{Shaw13} and \citet{Lan15} were featureless and could not confirm this result. The magnitudes of the source at the times these spectra were taken are unknown. Very recently, combining optical \citep{Shaw13} and gamma-ray results, the redshift has been constrained to be in the range 0.14 $\le z\le$ 0.53  at the 95 \% confidence limit \citep{Abda20}.
 We were able to obtain a very high S/N EFOSC2 spectrum (S/N $\ge$ 200, see Table \ref{tabres1}) of PKS 1440$-$389 (see Fig. \ref{fig_spec1}, second row, right). Careful inspection of the spectrum reveals the presence of CaHK, CaIG, and NaID features at $z$ = 0.1385  $\pm$ 0.0005.  We note that at that redshift the Mgb feature of the galaxy falls into the strong Galactic NaID absorption, and is thus undetectable. This result is consistent with the range obtained by \citet{Abda20} within slightly more than 2$\sigma,$ but it contradicts the tentative redshift published by 6dF galaxy survey.  

 The non-detection of these features in previous spectra is consistent with their low S/N  assuming a similar optical spectral state. The host galaxy magnitude is average: M$_R$ = -22.4 $\pm$ 0.2.

 \subsection{PMN J1457$-$4642}

 PMN J1457$-$4642 is one of the weakest and least studied of the gamma-ray sources in our sample, and no spectroscopic observation has been reported. Its EFOSC2 spectrum (see  Fig. \ref{fig_spec1}, third row, left) is clearly dominated by the emission of the host galaxy at  redshift $z$ = 0.1116 $\pm$ 0.0002, with a measurable component from a non-thermal power law and a weak (EW$\sim$ 5 \AA) H$\alpha$-[NII] emission complex. The rest-frame power law-to-galaxy ratio at 5500 \AA~is 0.3 $\pm$ 0.2. We note that the galaxy spectral shape cannot be well fitted with the local spectroscopic template of \citet{Man01}. A better though not completely satisfying fit can be achieved with a \citet{Bru03} template built with a 11 Gyr simple stellar population.

\subsection{NVSS J151148$-$051345}

\citet{Alv16} first reported a low S/N featureless optical spectrum for the HBL \citep{Tak13} NVSS J151148$-$051345. Recently, in an S/N $\sim$ 200, resolution $\lambda / \Delta \lambda \sim$ 250, GTC spectrum, a single feature around 4053 \AA~with EW=2.1 \AA~was detected \citep{Pai17b}. Interpreting this feature as an unresolved MgII doublet implies that the redshift of the source is greater than 0.45. In order to resolve this feature we observed the source with EFOSC2 using Grism 14, which allows for $\lambda / \Delta \lambda \sim$ 600 in the 3500-5000 \AA~wavelength range. 

We obtained a S/N=23 spectrum in which we detect a double-peaked spectral feature at the position discussed above with total EW 2.6$\pm$0.3 \AA~(see  Fig. \ref{fig_spec1}, third row, right).  We fit the feature with an MgII doublet using vpfit \citep{Cars14} obtaining a
$\chi^2_{\rm\,dof} \sim $1.07 for a redshift $z$ = 0.4480 $\pm$ 0.0003. We consider it a firm result, therefore, that the redshift of NVSS J151148$-$051345 is greater than 0.448. Finally the EWs of the two MgII components are about 1.7 and 0.9 \AA, their ratio is about 1.89, compatible  with optically thin gas \citep[see e.g.][] {Rag16}.

 \subsection{TXS 1515$-$273}

 TXS 1515$-$273 has  recently been recognised as a BL Lac \citep{Lef17}.  An upper limit to its redshift, $z\le$ 1.1 has been established from the lack of detection of the Ly $\alpha$ break in UVOT and SARA photometry \citep{Kau18}. Moreover, a low S/N spectrum taken by \citet{Pen17} with the Goodman Spectrograph at the SOAR telescope did not show any features, leaving its redshift unknown. 
  
  In February 2019 it was detected in VHE gamma rays with the MAGIC telescope, triggered by a high state in HE gamma rays \citep{Mir19}. Observations from the Tuorla blazar monitoring programme \citep{Nil18} show that during that time TXS 1515$-$273 was in a high state also in the optical band (around R$_c$=15.4), but by the time of our spectroscopic observations (June 2019) the flux had decreased significantly (to R$_c$=16.1){\footnote{\url{http://users.utu.fi/kani/1m/TXS_1515-273.html}}}. The Tuorla blazar monitoring magnitudes were derived with standard differential photometry and the comparison and control stars were calibrated using two nights of good weather data.
  We obtained a high S/N spectrum of TXS 1515$-$273 (see Fig. \ref{fig_spec1}, fourth row, left) in which absorption and emission lines at  $z \sim$ 0.1285 are visible along with a strong continuum. In absorption we detect with good confidence CaHK, CaIG, Mgb, and NaID. In emission [OII] and [OIII] $\lambda$ 5007 are detected  with good confidence, while the [NII]/H$\alpha$ complex is possibly present around 7410 \AA. The spectrum is very noisy at those wavelengths, however, due to instrumental effects; therefore, we do not  analyse this possible feature in detail. The fit of the detected features gives a precise redshift value $z$ = 0.1284 $\pm$ 0.0003.

  The magnitude of the spectrum after dereddening, telluric, and slit loss correction is R$_{\rm c}$=15.6 $\pm$ 0.1 (see Table \ref{tabres1}); the observed magnitude without corrections is  $R_{\rm c}$=16.2 $\pm$ 0.2, fully compatible with the results of the Tuorla blazar monitoring.  The host galaxy magnitude is M$_R$ = -22.4 $\pm$ 0.2. The rest-frame power law-to-galaxy ratio at 5500 \AA~is 3.0 $\pm$ 0.3. These results are broadly consistent with those presented by \citet{Bec20}.

 \subsection{NVSS J152048$-$034850}

NVSS J152048$-$034850 is a BL Lac object, possibly an LBL or an IBL \citep{Tak13}. \citet{Shaw13} report a low S/N detection of an MgII system at $z$ = 0.867 in a Palomar spectrum. Additionally, \citet{Kau17} report a redshift estimation of $z$ = 1.46 $\pm$ 0.1 from a photometric detection of the Ly $\alpha$ break using {\em Swift}/UVOT and Gamma-Ray burst Optical and Near-infrared (GROND) photometry.
In order to assess this lower limit, we observed this source with EFOSC2 employing Grism 8 with wavelength range 4500-6200 \AA~and resolution 7.4 \AA~achieving  S/N  S/N=43. We detect a clearly asymmetric spectral feature at the position reported by \citet{Shaw13} with total EW 3.65 $\pm$ 0.7 \AA.
Single component fits yielded unacceptable chi-square values and line widths up to three times the spectral resolution. We then fit the feature with an MgII doublet using vpfit \citep{Cars14} obtaining a $\chi^2_{\rm\,dof} \sim $1.0
for a redshift $z$=0.8680 $\pm$ 0.002 (see Fig. \ref{fig_spec1}, fourth row, right). Therefore, we consider it a firm result that the redshift of NVSS J152048$-$034850 is greater than 0.8680. The EWs of the two components of the fit are roughly 2.3 and 1.2 \AA. Their ratio is about 1.9, suggesting an optically thin gas \citep[see e.g.][] {Rag16}. Finally, we note that NVSS J152048$-$034850 is a high-redshift BL Lac object with a possible redshift of  z$\sim$ 1.46, as suggested by \citet{Kau17}, which makes it a very interesting target for future observations with CTA\footnote{ Photons at energy levels up to 100 GeV at $z$ = 1.5 and higher have been detected by {\em Fermi}-LAT; see section 3.6 and Figure 17 in \citet{Fer3FHL17}. CTA is expected to have a better sensitivity than {\em Fermi}-LAT at those energies.}
 
\begin{figure*}[htbp!]
   \centering
 \includegraphics[width=6.8truecm,height=5.75truecm]{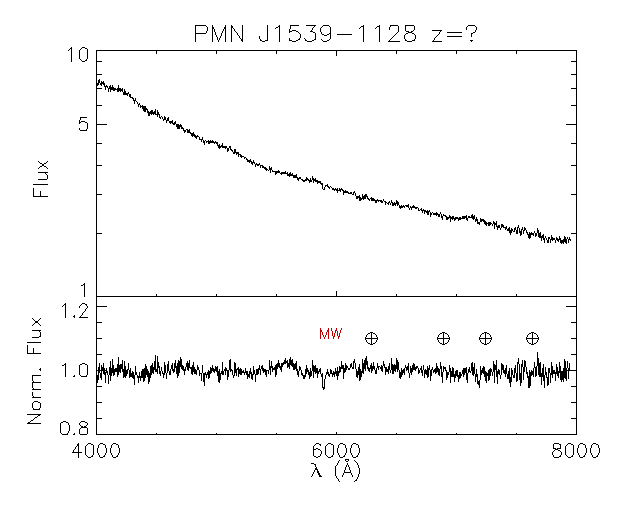}  \includegraphics[width=6.8truecm,height=5.75truecm]{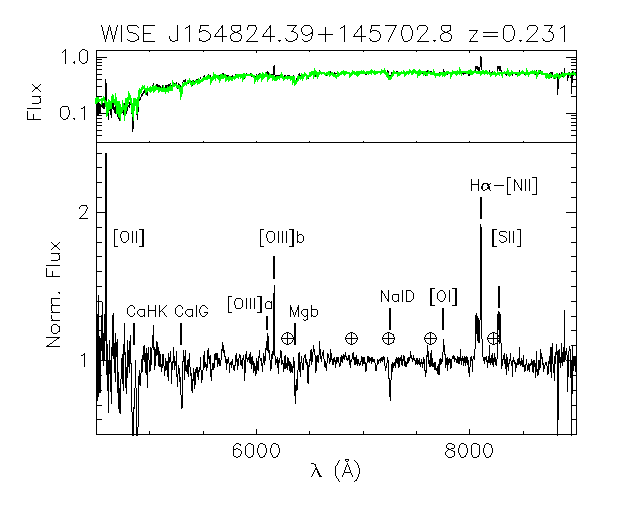}  
  \includegraphics[width=6.8truecm,height=5.75truecm]{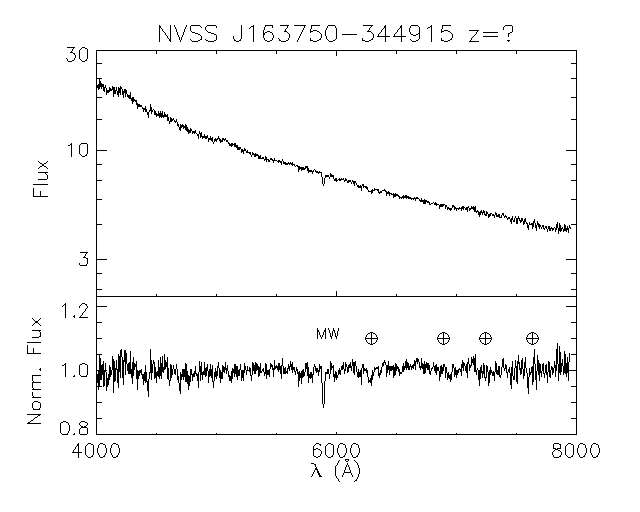}  \includegraphics[width=6.8truecm,height=5.75truecm]{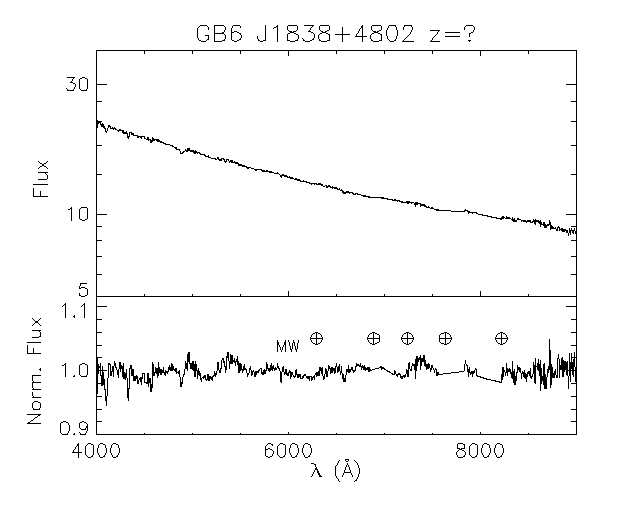}
 \includegraphics[width=6.8truecm,height=5.75truecm]{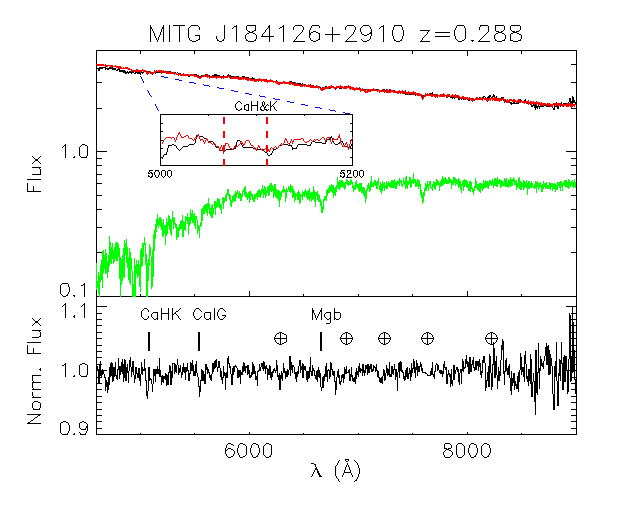}  \includegraphics[width=6.8truecm,height=5.75truecm]{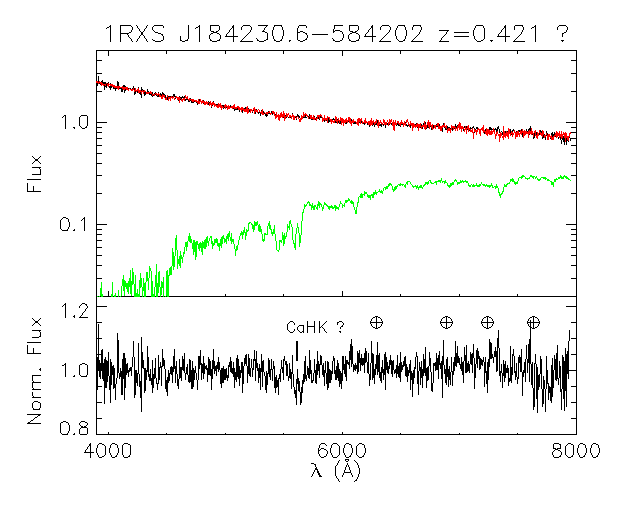}
  \includegraphics[width=6.8truecm,height=5.75truecm]{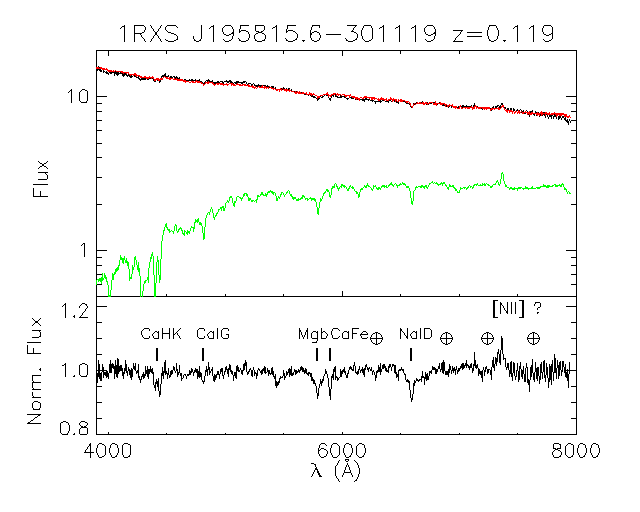}  \includegraphics[width=6.8truecm,height=5.75truecm]{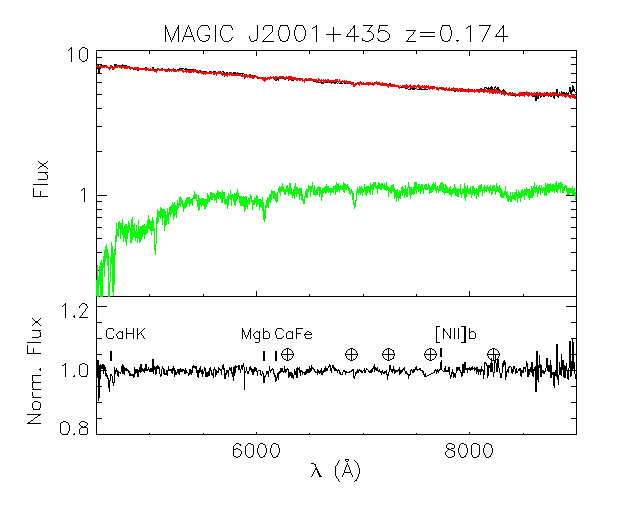}
   \caption{Same as Fig. \ref{fig_spec1} for sources 9 to 16 in Table \ref{tabobs1} }
\label{fig_spec2}
    \end{figure*}


\subsection{PMN J1539$-$1128}

  A spectrum of PMN J1539$-$1128 was reported by \citet{Pen17}. Using the Goodman spectrograph at the SOAR telescope, they obtained a featureless spectrum with S/N around 100. We observed the source with EFOSC2 using Grism 6 for 5400 sec, obtaining a spectrum with median S/N = 80 (see Fig. \ref{fig_spec2}, first row, left). With the exception of the Galactic NaID, no spectral feature was detected and the redshift of PMN J1539$-$1128 remains unknown.

 \subsection{WISE J154824.39$+$145702.8}

 The optical counterpart of WISE J154824.39$+$145702.8 is classified as a galaxy in the Sloan Digital Sky Survey (SDSS) \citep{Ahu19} with photometric redshift $z$ = 0.217$\pm0.02$. \citet{Alv16b} performed a 40-minute observation with the Device Optimized for the LOw RESoloution (DOLORES) spectrograph at the Telescopio Nazionale Galileo (TNG). The resulting 3700 - 8100 \AA~spectrum  is dominated by the host galaxy emission at a redshift of $z$ = 0.231. 
We took a one-hour spectrum of WISE J154824.39$+$145702.8 with ESI at Keck. We confirm the result of \citet{Alv16b} that the optical emission of the source is dominated by the host galaxy at $z \sim$ 0.231 (see Fig. \ref{fig_spec2}, first row, right). The rest-frame jet-to-galaxy ratio is lower than 0.07 (3$\sigma$ limit). The best fit is obtained using the  template by \citet{Man01}. Thanks to our higher sensitivity and wider wavelength range, we detect several additional emission lines: [OII], [OIII]$\lambda$\,4959, [OIII]$\lambda$\,5007, [OI]$\lambda$\,6300, H$\alpha$, [NII]$\lambda$\,6548, [NII]$\lambda$\,6583, and the [SII] doublet  (see Table \ref{WISEtab}).
The fit of the detected features gives a precise redshift value $z$=0.2308 $\pm$ 0.0002.
 Some lines have an EW  greater than 5 \AA, the standard BL Lac limit, showing that this host galaxy is peculiarly active and gas-rich for a BL Lac object. In order to investigate this issue, we measured their EWs by fitting Gaussian functions; the results are listed in Table \ref{WISEtab}. Computing the ratios [OIII]/[OII] and [OI]/H$\alpha$ and using Fig. 5 in \citet{Kew06}, the host galaxy can be classified as a LINER.

 The SED shown in \citet{Fuj16} shows the presence of a peak in the  NIR--optical domain. The above results show that it is due to the galaxy emission. The very low jet-to-host ratio suggests an HBL or possibly EHBL nature of the source, but the lack of multiwavelength data (in particular X-ray data) makes  a firm conclusion impossible at this moment. More observations are needed to settle this question.

\begin{table}
\caption{ Equivalent widths (in \AA)~of the emission lines detected in WISE J154824.39$+$145702.8.}
\begin{center}
\begin{tabular}{cc}
\hline
\hline
  Line &  Equivalent Width \\
          &     \AA     \\
   (1) &    (2)  \\          
\hline
    [OII]$\lambda$ 3727      &    18.8   $\pm$ 0.3 \\
    H$\beta$                 &    $\le$ 1           \\
 {[OIII]$\lambda$ 4959 }     &    1.6 $\pm$ 0.2      \\
  { [OIII]$\lambda$ 5007}    &    4.8 $\pm$ 0.2     \\
   {[OI]$\lambda$ 6300 }     &    2.2 $\pm$ 0.7     \\
   H$\alpha$                 &    4.5 $\pm$ 0.4     \\
   {[NII]$\lambda$ 6548}     &    4.4 $\pm$ 0.1     \\
   {[NII]$\lambda$ 6583 }     &    13.3 $\pm$ 0.1    \\
   {[SII]$\lambda$ 6716}     &    4.1 $\pm$ 0.3     \\
   {[SII]$\lambda$ 6731 }    &    4.4 $\pm$ 0.5     \\
\hline
\hline

\end{tabular}
\end{center}
\label{WISEtab}
\end{table}%

\subsection{NVSS J163750$-$344915}

 \citet{Pen17} reported on a spectrum of NVSS J163750$-$344915 obtained with the Goodman Spectrograph at SOAR (4000-7000 \AA). The spectrum has S/N = 60 and is featureless, confirming its BL Lac classification. We took an EFOSC2 spectrum of the object aiming to obtain a better S/N. However, we could only integrate for 45 min due to bad weather. The resulting spectrum has a S/N = 80 and is featureless (see Fig. \ref{fig_spec2}, second row, left). The redshift of NVSS J163750$-$344915 remains undetermined.

\subsection{GB6 J1838$+$4802}

No spectroscopic redshift measure is available for GB6 J1838$+$4802, but its redshift has been photometrically estimated at $z$ $\sim$ 0.3 from the detection of the host galaxy at magnitude R$_{\rm c}$=19.12 $\pm$ 0.05 in imaging observations \citep{Nil03}. Even if the redshift estimates depend on assumed host galaxy properties, and are therefore by definition less accurate than spectroscopic measurements, the detection of the host galaxy supports the possibility of also measuring the redshift  spectroscopically. However, our observation of GB6 J1838$+$4802 resulted in a featureless spectrum (Fig. \ref{fig_spec2}, second row, right), despite its very high S/N ($\sim$ 250). We note that this non-detection may be due to the brightness of the source; at R$_{\rm c} \sim$ 15.8 the nuclear emission is about 3.3 magnitudes (about 20 times) stronger than the host emission. In a case like this one the best option may be to trigger another spectroscopic observation during a minimum in the optical light curve.

\begin{figure*}[h!]
   \centering
   \includegraphics[width=6.8truecm,height=5.75truecm]{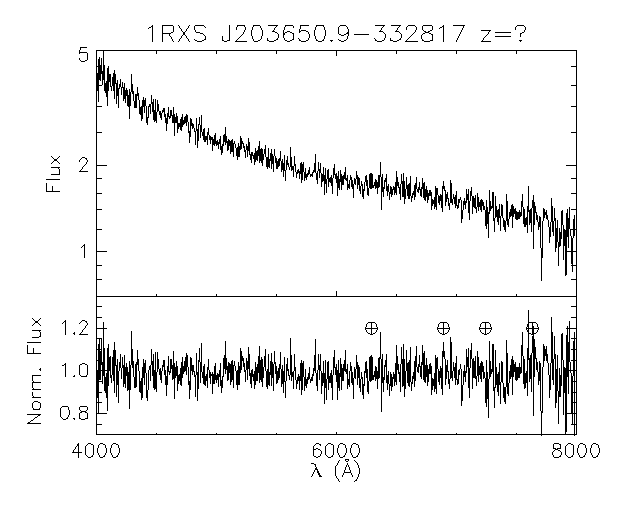}  \includegraphics[width=6.8truecm,height=5.75truecm]{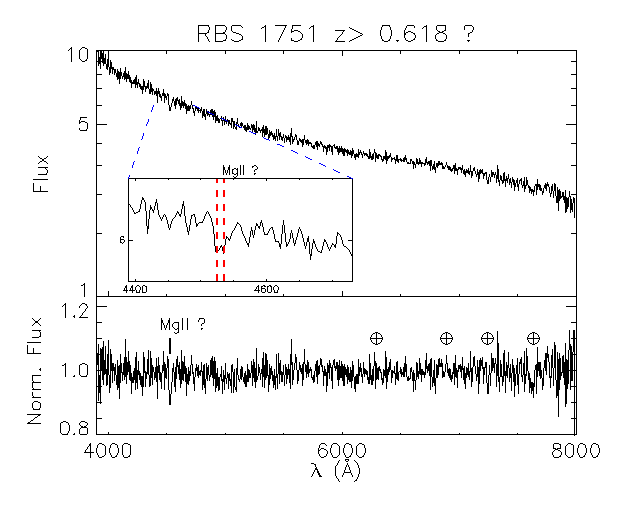}
\includegraphics[width=6.8truecm,height=5.75truecm]{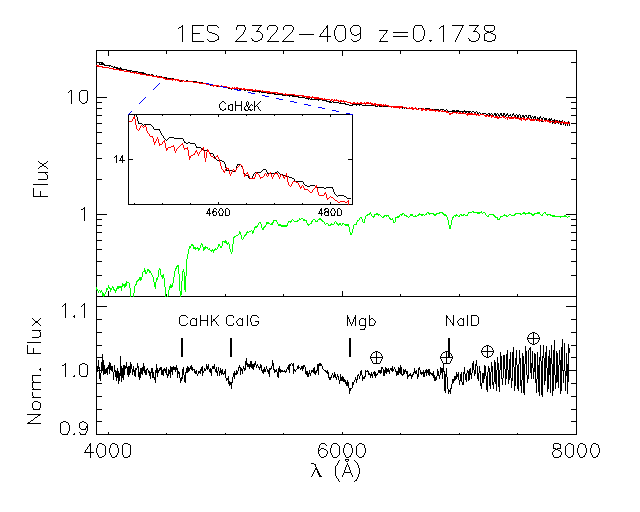}

   \caption{Same as Fig. \ref{fig_spec1} for the last three sources in Table \ref{tabobs1}}
\label{fig_spec3}
 \end{figure*}

\subsection{MITG J184126$+$2910}

The optical counterpart of MITG J184126$+$2910 was identified by \citet{Marchesini16}. The moderate S/N, featureless, optical spectrum they took with the DOLORES spectrograph at the TNG confirmed its classification as a BL Lac object. A later spectrum \citep{Marchesi18} taken with the Kitt Peak Ohio-State Multi-Object Spectrograph (KOSMOS) spectrograph at the Mayall Telescope is also featureless. 

We note that the infrared counterpart, 2MASX J18412170$+$2909404, is extended, suggesting that the emission of the host galaxy may be detectable. Our Keck/ESI spectrum (see Fig. \ref{fig_spec2}, third row, left) reaches an average S/N  of 100. In the mostly featureless spectrum we  identified the CaHK doublet and CaIG features at 5$\sigma$ and 4$\sigma$, respectively, at  redshift $z \sim$ 0.288. We also possibly identify  the Mgb and CaFe features at about 3-4$\sigma$ at the same redshift. A precise redshift $z$ = 0.2883 $\pm$ 0.0003 is obtained fitting these features. 
The absolute magnitude of the host galaxy is M$_{\rm R}$=$-$22.9 $\pm$ 0.3, bright but within the expected range.

\subsection{1RXS J184230.6$-$584202}

Two spectra of 1RXS J184230.6$-$584202 have been reported recently. One spectrum \citep{Des19} was taken on June 1, 2018, with KOSMOS at the 4m Mayall telescope. The source had magnitude V=17.5, and the spectrum was featureless with a S/N of 30. The other was  taken with the Goodman spectrograph on the SOAR telescope \citep{Marchesini19} on May 9, 2017. The S/N is similar to the first one, but in this case the authors report the detection of the CaHK and CaIG features at  redshift $z$ = 0.421. The magnitude of the source is not reported by these authors.
 
 In order to verify the presence of these features, we observed 1RXS J184230.6$-$584202 for 1 hour and 15 minutes obtaining a S/N of 35.  The spectrum has a power-law shape with magnitudes V=18.8 $\pm$ 0.2 and R$_{\rm  c}$=18.3 $\pm$ 0.2  (see Fig. \ref{fig_spec2}, third row, right). After careful examination we possibly detect at the 3.6 $\sigma$ level the CaHK feature at $z \sim$ 0.421 \citep[the same redshift as in][]{Marchesini19}. We note that at  magnitude V = 18.8 the source was much fainter than in \citet{Des19},  which may explain our weak detection and their non-detection. Considering the low significance of the detection, we consider this redshift value as tentative.

\subsection{1RXSJ195815.6$-$301119}

  Two independent publications give the same value for its redshift, $z$ = 0.119. This value is reported in the DR3 6dF catalogue \citep{Jon04,Jon09} and in the Reflex Cluster survey \citep{Guz09}. The public 6dF spectrum has a very low S/N (see Fig. \ref{fig_newplots}), while the Reflex spectrum is not available.
  
  We thus took a 30 min spectrum with EFOSC2 in order to confirm or disprove this result. The resulting high S/N spectrum is a combination of a power law with the emission of an elliptical galaxy modelled with an 11 Gyr template \citep{Bru03} at $z$ = 0.1190 $\pm$ 0.0003.  Weak [NII] emission is also detected (see Fig. \ref{fig_spec2}, fourth row, left). The rest-frame power law-to-galaxy ratio at 5500 \AA~is 1.2 $\pm$ 0.2 and the galaxy has absolute magnitude M$_{\rm R}$=-22.1 $\pm$ 0.2. This is thus a firm redshift value.

 \subsection{MAGIC J2001$+$435}

  MAGIC J2001$+$435 is a BL Lac object \citep{Bas09} detected at TeV energies \citep{Ale14b}. The host galaxy has been detected in deep imaging secured during low blazar activity. The host magnitude was measured to be I=17.15 $\pm$ 0.06, which was translated into a photometric redshift $z$ = 0.18 $\pm$ 0.04 \citep{Ale14b}.  A high S/N ($\sim$ 150-200) optical Keck/Low Resolution Imaging Spectrometer (LRIS) spectrum was taken by \citet{Shaw13} on August 18, 2009, but no spectral features could be detected. From the spectrum (R. Romani, private communication) the source had an observed magnitude R$_{\rm c}$=14.8, more than 15 times stronger than the host.

  In May 2018  the Tuorla blazar monitoring light curve\footnote{{\url{http://users.utu.fi/kani/1m/MG4_J200112+4352.html}}} showed that MAGIC J2001+435 was in a low state with an observed magnitude more than two magnitudes fainter than at the epoch of the Keck/LRIS spectrum. It was decided, therefore, to observe it with Keck/ESI in place of a previously scheduled target.
  
  The exposure time was 3480 sec and the resulting S/N  was 105. In the final spectrum we estimate a redshift $z$=0.1739 $\pm$ 0.0002 from the detection of the CaHK, Mgb, CaFe, and possibly NaID absorption features (Fig. \ref{fig_spec2}, fourth row, right). At a 4.6$\sigma$ level we also detect the [NII]$\lambda$ 6583 emission line at the same redshift. The result is consistent with the imaging estimate in \citet{Ale14b} at less than 1$\sigma$. 
    Our redshift measurement was possible even though the S/N of our spectrum is much lower than that  from \citet{Shaw13}, demonstrating that waiting for an optical low state may be very effective for blazar redshift measurement.

\subsection{1RXS J203650.9$-$332817}

1RXS J203650.9$-$332817 was observed by \citet{Alv16b} who took a 20-minute spectrum with the Goodman spectrograph on the SOAR telescope. The spectrum has a S/N of about 50 and a  general power-law shape consistent with  classification as a BL Lac object. A weak feature around 5000 \AA, if interpreted as CaHK, sets a redshift value of $z$ = 0.237.

 Due to the weakness of the feature we observed the source with EFOSC2 in order to confirm or disprove this result with a higher S/N spectrum. However, due to bad weather, we were only able to obtain a low S/N featureless spectrum (Fig. \ref{fig_spec3}, first row, left). The redshift of 1RXS J203650.9$-$332817 remains undetermined.

\subsection{RBS 1751}

RBS 1751 was   classified as a BL Lac object on the basis of a noisy spectrum from the 6dF survey \citep{Jon04, Jon09}. We observed the source with EFOSC2 for one hour in difficult atmospheric conditions. The resulting spectrum has a S/N of 40, higher than the 6dF value, and has a power-law shape (Fig. \ref{fig_spec3}, first row, right).  Only after normalisation can a weak absorption feature with EW = 2.3 $\pm$ 0.5 \AA~centred around $\sim$ 4530 \AA~  be found. If we interpret this feature as being due to an unresolved MgII system, then the source would be located at $z \ge$ 0.618. Unfortunately, the grism we used, Grism 6, does not have enough spectral resolution to separate the two components of a possible MgII absorber. Interestingly, the only possible feature present in the 6dF spectrum is at the same wavelength; however, it is very weak. A deeper observation with better spectral resolution and S/N is needed to solve this issue. We conclude that RBS 1751 is tentatively at a redshift $z \ge$ 0.618. We provide no error estimate on this value as the possible feature could not be fitted.

\subsection{1ES 2322$-$409}

1ES 2322$-$409  is a bright TeV emitter, serendipitously discovered by  H.E.S.S. \citep[see][]{HESS19}. The redshift of this source is uncertain; a value $z$ = 0.174 has been suggested on the basis of a low S/N 6dF spectrum \citep{Jon04, Jon09}. We note that the redshift measurement was mostly based on a weak feature at $\sim$ 6900 \AA, in a region where there is  strong telluric absorption.  A higher S/N spectrum taken with the FOcal Reducer and low dispersion Spectrograph (FORS) at the Very Large Telescope was featureless \citep{Lan13}. The spectrum is available on the ZBLLAC\footnote{http://web.oapd.inaf.it/zbllac/} database and is quite noisy. When the FORS spectrum was taken the source was at magnitude R= 15.7 \citep{Lan13};  there is no information on the source magnitude during the 6dF observation. 
  
   In order to clarify this issue we took a high S/N EFOSC2 spectrum (Fig. \ref{fig_spec3}, second row). The source was at magnitude R$_{\rm c}$=16.1 and the spectrum is dominated by a non-thermal continuum. In the continuum we were able to detect several weak features: CaHK, CaIG, Mgb, and NaID at a common redshift $z$ = 0.1736 $\pm$ 0.0008. We consider this a firm redshift. We thus confirm the 6dF result, despite the low S/N of their spectrum. The detection of the NaID feature may have been possible due to the source being in a low state.

  \section{Discussion and conclusions}

We observed 19 BL Lacs  detected at E$\ge$ 10 GeV with the {\em Fermi}-LAT satellite and/or by ground-based Cherenkov telescopes. 
The observations were performed with the ESI spectrograph at the Keck Observatory,  with the RSS spectrograph at the South African Astronomical Observatory and with the EFOSC2 spectrograph at the ESO/La Silla observatory. The observing strategy called for obtaining, whenever possible, S/N  values per pixel greater than 100. We aimed to measure or constrain the redshift and measure the properties of the host galaxy of each of our target BL Lacs. In the following subsections we discuss these results in more detail and conclude with a summary.

 \subsection{Spectral signal-to-noise ratio}
 
  Only nine of our targets reached the target S/N:  1RXS J015658.6$-$530208, 1RXS J020922.2$-$522920, PKS 1440$-$389, TXS 1515$-$273, GB6 J1838$+$4802, MITG J184126$+$2910, 1RXS J195815.6$-$301119, MAGIC J2001$+$435, and 1ES 2322$-$409. We measured the redshift for eight of them.
  The other ten have S/N values between 20 and 80, and from them we could measure only three firm redshifts and one tentative value. In addition we measured two firm lower limits and one tentative lower limit.  
  
   A high S/N spectrum thus appears to be necessary to allow  the measurement of the redshift for the gamma-ray BL Lacs in our sample. On the other hand, a low S/N spectrum, while useful for classification, is not likely to succeed in this effort. The comparative success of high S/N spectra with respect to those with low S/N supports our strategy.  
  However, we also note that in the case of GB6 J1838$+$4802, even a very high S/N spectrum did not lead to a successful measurement of the redshift. This is in line with several previous works \citep{Pai17a,Pit14} where it was not possible to measure redshifts of well-known blazars, even with very high-quality spectra. Therefore, while for gamma-ray BL Lacs reaching a S/N of at least 100 is a good indicator of the success of a redshift measurement, it is not a guarantee.
 Additionally, both of our SALT/RSS spectra and three of  our four Keck/ESI spectra reached our target S/N,  while only 4 out of 13 EFOSC2 spectra reached that goal. While the sample is still small, this suggests that 10m class telescopes are the best tools for our programme. 
    Finally, technical or weather issues  prevented us from reaching the target S/N  in ten of our spectra; three of them have no redshift measurement or lower limit. We will re-observe these three sources to secure a spectrum with a S/N  of 100. For any sources for which the redshift will not have  beeen measured even then, we plan to organise dedicated target of opportunity programmes to obtain high S/N spectroscopy during their optical low states.

 \subsection{Optical extension of the sources}
 
 We had eight targets with extended counterparts:  four of them have extended NIR counterparts in the 2MASS Extended Source Catalogue (2MASX) \citep{Jar00}. Of the remaining four,  the extension has been detected for three of them in dedicated observations  and the last one, WISE J154824.39$+$145702.8, is classified as extended in the SDSS database.
   The 2MASX sources are TXS 1515$-$273, PMN J1457$-$4642, MITG J184126$+$2910, and 1RXS J195815.6$-$301119. We measured the redshift for all of them with values between 0.1116 and 0.288 and an average value  <$z$(2MASX)> = 0.162. This may be an indication that gamma-ray BL Lacs whose counterparts belong to the 2MASX catalogue are good candidates for redshift measurement provided a sufficient S/N is obtained. The low average value of the redshift may be due to the relatively low sensitivity of the 2MASX survey that can reliably detect  the extension only for sources with  J $\ge$ 15.0\footnote{https://old.ipac.caltech.edu/2mass/releases/allsky/doc/sec6$\_$1h.html}, which is the case for the four sources mentioned above.

  The remaining extended sources are GB6 J1838$+$4802 \citep{Nil03}, WISE J154824.39$+$145702.8, PKS 1440$-$389 (Fallah Ramazani et al. in preparation), and MAGIC J2001$+$435 \citep{Ale14b}. The redshifts of the last three were measured. As discussed above, our observations of PKS 1440$-$389 are much deeper than previous observations, but no information on its optical state during previous observations is available, and we confirmed the previous redshift measurement of WISE J154824.39$+$145702.8.
  The situation is different for the other two sources; MAGIC J2001$+$435 was in an optical low state, which eased the redshift measurement, while a previous more sensitive observation failed to detect any features. Conversely GB6 J1838$+$4802 was in an optically high state, which likely hampered the redshift measurement.
  Overall, we measured seven redshifts for eight extended targets, which confirms that previous results \citep{Nil03} of the $\sim$ 90 \% efficiency of redshift measurements for extended counterparts. Moreover, the cases of GB6 J1838$+$4802 and MAGIC J2001$+$435 show that the optical state of the BL Lac object can be relevant for our purpose, and that using the optical state of the BL Lac object as a further parameter of our strategy may improve the efficiency of our programme.

  \subsection{Properties of the host galaxies}

   The average magnitude of the 11 firmly detected host galaxies is M$_\mathrm{R}$ = -22.6 $\pm$ 0.4, slightly brighter than but compatible with the value reported in \citet{Shaw13}, and fainter than the values in \citet{Sbar05} and \citet{Pit14}. The best fit template was the local one \citep{Man01} in nine cases. In the remaining cases the best fit was produced by a simple stellar population template \citep{Bru03} with age 11 Gyr (PMN J1457$-$4642 and 1RXS J195815.6$-$301119) and 2.5 Gyr  (1RXS J011501.3$-$340008). This is broadly in agreement with previous results that host galaxies of BL Lacs are normal elliptical galaxies \citep{Urr00}.
   We detected faint and narrow emission lines in only five of our targets, in line with previous results at similar sensitivities \citep{Pit14}.
    We did not estimate upper limits on the magnitudes of the non-detected host galaxies because the S/N of their spectra  was not high enough to yield a meaningful limit.

\subsection{Comparison with other campaigns}

We cannot present  an exhaustive literature review here, but  we briefly compare the first results from our programme 
to those of the main previous and ongoing spectroscopic observations aimed at measuring the redshifts of gamma-ray blazars. While we present here observations of 19 sources, our campaign targets 165 sources; we thus limit our comparison to campaigns targeting extensive samples with about 100 targets or more: the  \citet{Shaw13} campaign, the  campaign described by \citet{Pena20}, and the  recently undertaken campaign by \citet{Pai20}.   With respect to the first one, our sample contains brighter sources with harder gamma-ray spectra,  but it contains fewer targets, while our spectra  have on average higher and more uniform S/N. Regarding the campaign by \citet{Pena20}, we target  brighter objects with harder
gamma-ray spectra and with  better determined classifications; we have fewer targets and we have much higher S/N spectra. With respect to both of these campaigns we can thus expect to have a higher redshift detection efficiency.
Finally, the campaign started by \citet{Pai20} includes 91 bright objects from the 3FHL catalogue and the authors plan to take high S/N spectra for all of them from the GTC telescope in the Canary Islands. Their sample is smaller than  ours and is limited to the Northern Hemisphere, and 
its gamma-ray selection is much simpler than ours.  Their selection is based on simple extrapolations of the flux and spectral index provided by the catalogue, while ours relies on full Monte Carlo simulations of the spectra and EBL absorption using the CTA instrumental response. 
We think we will have a similar redshift detection efficiency on a larger sample extended to the Southern Hemisphere. 
With respect to all of these campaigns, we will also add imaging information which will deepen our knowledge of the 
targets, for example by  allowing us to determine (or constrain) more precisely the magnitude of the host galaxy.

  \subsection{Summary}

We report on the first results of our programme aimed at measuring the redshift of gamma-ray bright blazars likely to be detected with CTA. Our main results are the following:
\begin{enumerate}
    \item  We performed spectroscopic observations of 19 gamma-ray blazars. Seventeen of them were previously observed in spectroscopy and seven uncertain redshift values were proposed in the literature; we confirmed four and disproved two of them, while the remaining
    one is still dubious. We measured 11 firm redshifts and 1 tentative redshift with values ranging from $z$ = 0.1116 to $z$ = 0.4824.  
    
     \item Six of the blazars are at redshifts of z > 0.2, where fewer than 15 VHE BL Lacs are currently known. In particular, we measured firm spectroscopic redshifts of three known TeV sources: PKS 1440-389, 1ES 2322-409, and MAGIC J2001+435. After our observations the number of TeV BL Lacs without redshift values decreases from 13 to 10.

  \item We confirmed two previously suggested spectroscopic lower limits and detected  one tentative lower limit with values ranging from $z \ge$ 
 0.4480 to $z \ge $ 0.8680.
 
    \item We achieved high efficiency (eight out of nine) in redshift measurement for high S/N spectra, and low efficiency (three-four out of ten) in redshift measurement for low S/N spectra.

    \item We achieved high efficiency (seven out of eight) in redshift measurement for BL Lacs with extended optical counterparts.
    
    \item We measured  the redshift of MAGIC J2001$+$435 during an optical low state. This is an indication that spectroscopic observations triggered during a spectroscopic minimum can improve redshift measurement efficiency.
 
    \item We measured  the average magnitude of the host galaxies, M$_\mathrm{R}$ = -22.6 $\pm$ 0.4, which is compatible with previous measurements of this quantity for gamma-ray blazars. The properties of the hosts are consistent with normal elliptical galaxies.
    
\end{enumerate}
 
   The overall redshift detection efficiency of  58\% or (11/19) is higher than that in the most complete survey of gamma-ray blazars  \citep{Shaw13} of 44 \%. This preliminary difference may be due to the different sample selection criteria and/or to the higher S/N  of our spectra on average. Due to bad weather and, in some cases, limited sensitivity, not all our spectra matched our criteria. 
   
   Our programme of redshift observations will continue on the objects from our sample with the goal of obtaining a spectrum with S/N  100 for each object, in an optical low state if needed. This will help to shape the CTA Key Science Programme on AGN of blazar type.  This programme will also serve the  astronomical community at large.  On the one hand, it will enable scientists studying blazar emission and populations to better tune their models. On the other hand, scientists applying for time not only with CTA, but also with present-day VHE observatories and with other multiwavelength facilities such as {\em Fermi}-LAT, {\em XMM-Newton}, {\em Chandra}, ALMA, and  VLA will be able to select optimal targets for their observing programmes from a larger pool of blazars with known redshifts.

\begin{acknowledgements}

 This paper went through internal review by the CTA consortium. We thank D. Horan and M. Errando who acted as internal CTA reviewers for comments that helped improve the quality of the paper. We thank an anonymous referee for his/her comments. We thank Elisabete de Gouveia del Pino for useful discussions. This research has made use of the CTA instrument response functions provided by the CTA Consortium and Observatory, see \url{https://www.cta-observatory.org/science/cta-performance/}(version prod3b-v1) for more details. We gratefully acknowledge financial support from the agencies and organisations listed here: \url{http://www.cta-observatory.org/consortium_acknowledgments} and in particular from the US National Science Foundation, award PHY-1707454.
The authors wish to recognise and acknowledge the very significant cultural role and reverence that the summit of Maunakea has always had within the indigenous Hawaiian community.  We are most fortunate to have the opportunity to conduct observations from this mountain.
Some of the observations reported in this paper were obtained with the Southern African Large Telescope (SALT).
Funding for SDSS-III has been provided by the Alfred P. Sloan Foundation, the Participating Institutions, the National Science Foundation, and the U.S. Department of Energy Office of Science. The SDSS-III web site is http://www.sdss3.org/. SDSS-III is managed by the Astrophysical Research Consortium for the Participating Institutions of the SDSS-III Collaboration including the University of Arizona, the Brazilian Participation Group, Brookhaven National Laboratory, University of Cambridge, Carnegie Mellon University, University of Florida, the French Participation Group, the German Participation Group, Harvard University, the Instituto de Astrofisica de Canarias, the Michigan State/Notre Dame/JINA Participation Group, Johns Hopkins University, Lawrence Berkeley National Laboratory, Max Planck Institute for Astrophysics, Max Planck Institute for Extraterrestrial Physics, New Mexico State University, New York University, Ohio State University, Pennsylvania State University, University of Portsmouth, Princeton University, the Spanish Participation Group, University of Tokyo, University of Utah, Vanderbilt University, University of Virginia, University of Washington, and Yale University.
W.M. acknowledges support from ANID projects Basal AFB-170002, PAI79160080 and FONDECYT 11190853.
This research has made use of the SIMBAD database, operated at CDS, Strasbourg, France.

 \end{acknowledgements}

\bibliographystyle{aa} 
\bibliography{ctaredshift}        

\begin{appendix}

\onecolumn
\section{Redshift values in our simulation sample modified with respect to the 3FHL}

\begin{table*}[!h]
\caption{\label{tabapp} Sources in our sample for which we modified the redshift value with respect to the 3FHL catalogue. The columns contain the source name, the redshift value in the 3FHL catalogue, the reference for that redshift value, the lower limit value, the adopted redshift value, the reference  justifying the adopted value, and finally, for sources discussed in this paper, the measured redshift.  If the entry (redshift, lower limit, reference) is unknown, the legend is `---'. There is a total of 32 sources with 7 new redshifts and 14 lower limits. The lower limit of PMN J0816$-$1311 ($z \ge$ 0.288) is smaller than the adopted value for unknown redshift $z$=0.3; we therefore used  $z$=0.3 for this source in our simulations. The spectra of ten sources have not been published yet to our knowledge, and we present them for reference in Fig. \ref{fig_newplots}. }
\centering
\begin{tabular}{|l|c|c|c|c|c|c|c|}
\hline
                 \multicolumn{3}{|c|}{3FHL}  &  \multicolumn{3}{|c|}{Simulations}     &     \multicolumn{2}{|c|}{Measured} \\
\hline
3FHL  name  & Source  name    &             &     redshift   & redshift     &  adopted    & adopted redshift/     &  
 \\
                       &                        & redshift  &   reference     & lower limit   &  redshift & lower limit reference & redshift \\
\hline
3FHL J0022.0$+$0006  & 1RXS J002200.9$+$000659  & ---          &   ---           & ---             & 0.3057      & (1)               &  --- \\
3FHL J0033.5$-$1921   & KUV 00311$-$1938                 & 0.61      &   (2)         & 0.506         & 0.506        & (3)              &  --- \\
3FHL J0120.4$-$2701   & PKS 0118$-$272                     & ---         &   ---           & 0.558        & 0.558        & (5)               &  --- \\
3FHL J0211.2$+$1051  & MG1 J021114$+$1051           & 0.2         &   (6)          & ---              & 0.3            & ---               &  --- \\
3FHL J0237.6$-$3602  & RBS 334                                 & ---          &   ---            & ---              & 0.411        & (3)              &  --- \\
3FHL J0338.9$-$2848  & NVSS J033859$-$284619     & ---           &   ---           & ---               & 0.251        & (4)             &  --- \\
3FHL J0449.4$-$4350  & PKS 0447$-$439                   & 0.205      &   (7)          & ---              & 0.3            & (3),(8)         &  --- \\
3FHL J0508.0$+$6737 & 1ES 0502$+$675                  & 0.416      &   (9)           & ---              & 0.3            & (10),(11)     &  --- \\3FHL J0521.7$+$2112 & TXS 0518$+$211                  & 0.108       &   (12)         & ---             & 0.3            & (8)               &  --- \\
3FHL J0550.5$-$3215  & PKS 0548$-$322                  & ---            &   ---            & ---             & 0.069         & (13)            &  --- \\
3FHL J0612.8$+$4122 & B3 0609$+$413                     & ---            &   ---           & 1.108        & 1.108         & (12)            &  --- \\
3FHL J0622.4$-$2606  & PMN J0622$-$2605              & 0.414       &  (14)         & ---             & 0.3             &  ---              &  --- \\
3FHL J0650.7$+$2503  & 1ES 0647$+$250                 & 0.203       &  (15)          & ---            & 0.3            & (8)               &  --- \\
3FHL J0816.4$-$1311  & PMN J0816$-$1311              & 0.046       &  (14)         & 0.288        & 0.3            & (3)               &  --- \\
3FHL J1107.4$+$0221 & NVSS J110735$+$022225   & ---             &  ---            & 1.0743      & 1.0743     & (1)               &  --- \\
3FHL J1120.8$+$4212 & RBS 0970                             & 0.124       &  (16)          & ---             & 0.3           & (8)               &  --- \\
3FHL J1253.1$+$5300 & S4 1250$+$53                      & ---           &  ---             & 0.664         & 0.664       & (1)               &  --- \\
3FHL J1410.5$+$1438 & NVSS J141028$+$143841  & ---            &  ---             & ---              & 0.144       & (1)               &  --- \\
3FHL J1427.0$+$2348 & PKS 1424$+$240                 & ---           & ---              & 0.604         & 0.604       & (17)             &  --- \\3FHL J1436.9$+$5639 & RBS 1409                             & 0.15       & (18)            & ---              & 0.3           & (12)             &  --- \\
3FHL J1442.5$-$4621 & SUMSS J144236$-$462302 &  ---          & ---              & ---               & 0.1026     & (14)             &  --- \\
3FHL J1443.9$-$3908 & PKS 1440$-$389                  & 0.069      & (14)           & ---               & 0.3           & (12)             &  0.1385 \\
3FHL J1447.9$+$3608 & RBS 1432                            &  ---          &  ---             & 0.739          & 0.739      & (1)               &  --- \\
3FHL J1603.8$-$4903 & PMN J1603$-$4904             &  ---           &  ---            & ---               & 0.231       & (19)              &  --- \\
3FHL J1610.6$-$6649  & PMN J1610$-$6649            &  ---           & ---             & 0.447          & 0.447      & (1)                &  --- \\3FHL J1903.2$+$5540  & TXS 1902$+$556               &  ---           & ---             & 0.727          & 0.727      & (12)              &  --- \\3FHL J1918.2$-$4111   & PMN J1918$-$4111             &  ---          & ---              & 1.591          & 1.591      & (12).            &  --- \\3FHL J1931.1$+$0937 & RX J1931.1$+$0937            & ---           & ---              & 0.476          & 0.476       & (12)            &  --- \\
3FHL J1958.3$-$3011  & 1RXS J195815.6$-$301119  & 0.119     & (14)           & ---                & 0.3           & ---              &  0.119 \\
3FHL J2243.9$+$2020 & RGB J2243$+$203              & ---           &  ---             & 0.395          & 0.395        & (12)          &  --- \\
3FHL J2255.2$+$2410 & MITG J225517$+$2409       & ---           &  ---             & 0.864          & 0.864       & (1)            &  --- \\
3FHL J2324.7$-$4040 & 1ES 2322$-$409                  & 0.174      & (14)           & ----               & 0.3           & ---             &  0.1738 \\

\hline

\end{tabular}
\tablefoot{(1) \citet{Bla17}; (2) \citet{Pir07}; (3) \citet{Pit14}; (4) \citet{Hal97}; (5) \citet{Vla97}; (6) \citet{Mei10}, imaging redshift; (7) \citet{Per98}, weak features, unconfirmed; (8)  \citet{Pai17}; (9) \citet{Landt02}; (10) \citet{Sca99}; (11) \citet{Gio04}; (12) \citet{Shaw13}; (13) \citet{Fos76}; (14) \citet{Jon04, Jon09}, low S/N spectrum; (15) \citet{Rec03}, tentative value;  (16) \citet{Per96}; (17) \citet{Fur13}; (18) \citet{Schwope00}; (19) \citet{Gol16}.}
\end{table*}%

\subsection{Notes on each source in Table \ref{tabapp}}
\begin{figure*}[htbp]
\begin{center}

 \includegraphics[width=0.45\hsize,height=0.20\hsize]{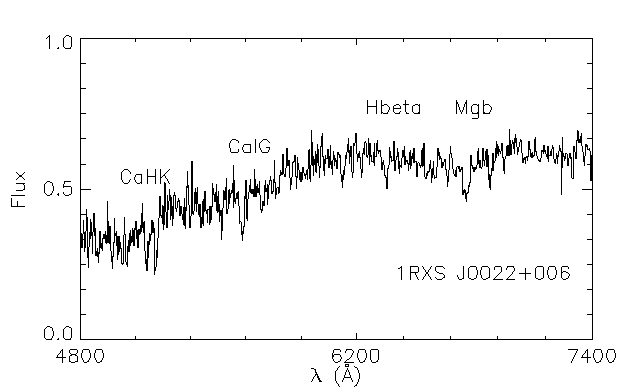} \includegraphics[width=0.45\hsize,height=0.20\hsize]{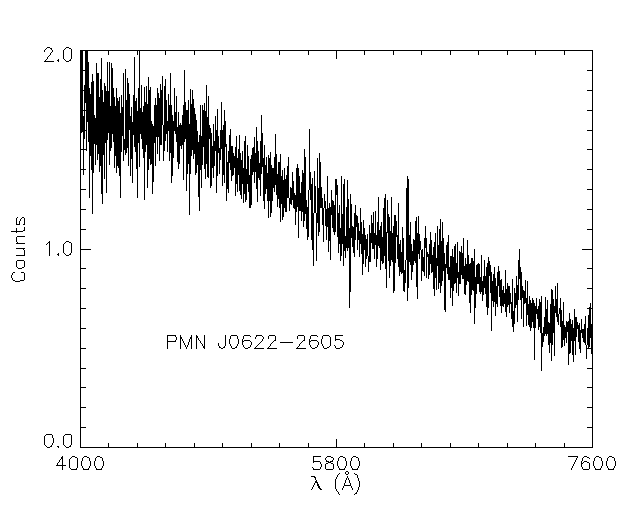} \includegraphics[width=0.45\hsize,height=0.20\hsize]{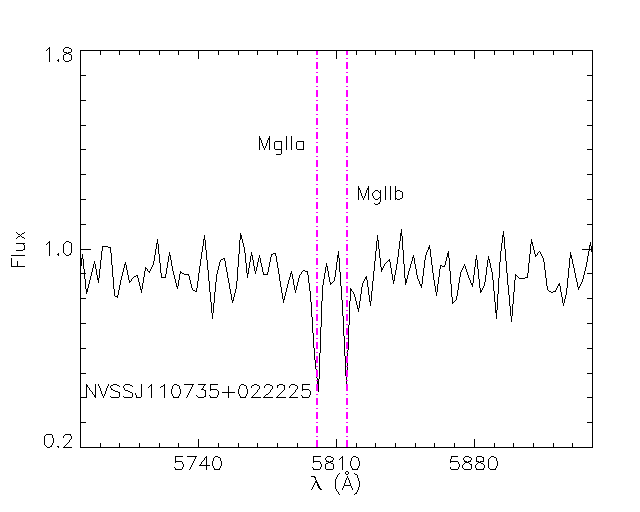} \includegraphics[width=0.45\hsize,height=0.20\hsize]{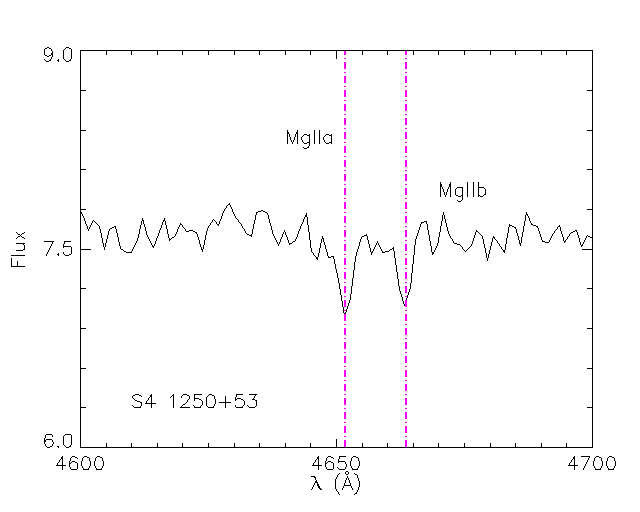}  \includegraphics[width=0.45\hsize,height=0.20\hsize]{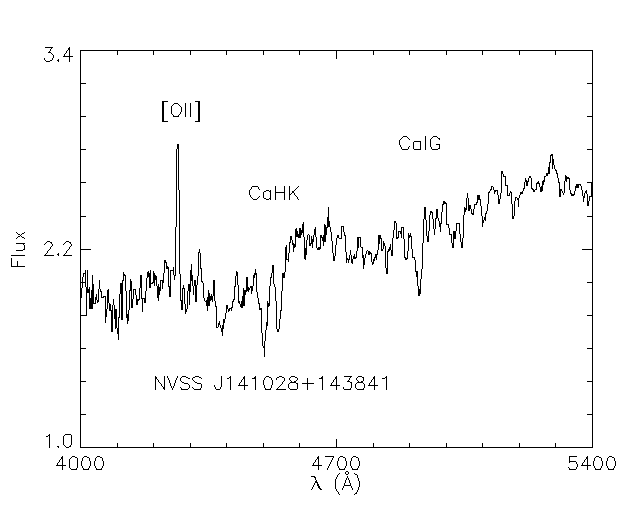} \includegraphics[width=0.45\hsize,height=0.20\hsize]{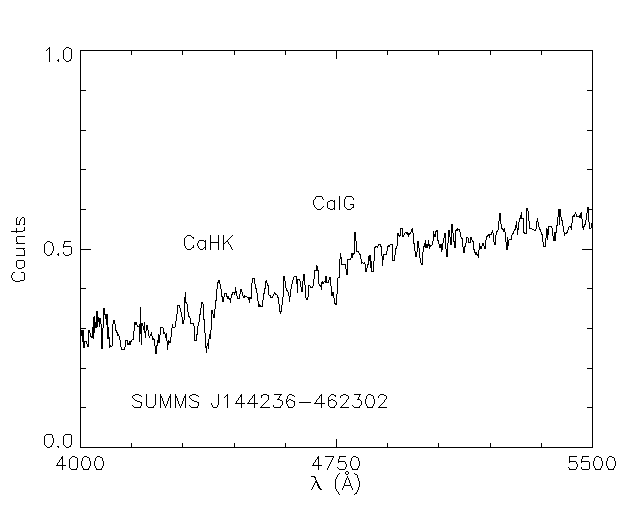}
\includegraphics[width=0.45\hsize,height=0.20\hsize]{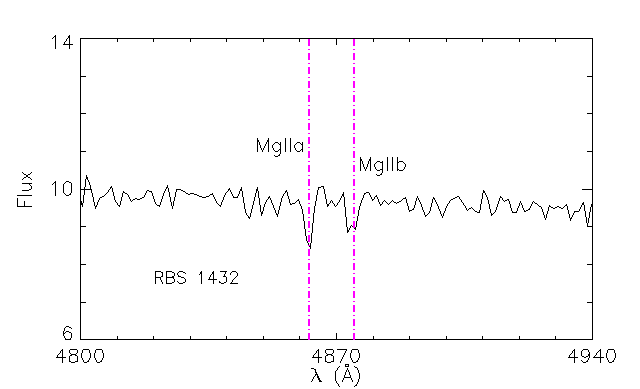} \includegraphics[width=0.45\hsize,height=0.20\hsize]{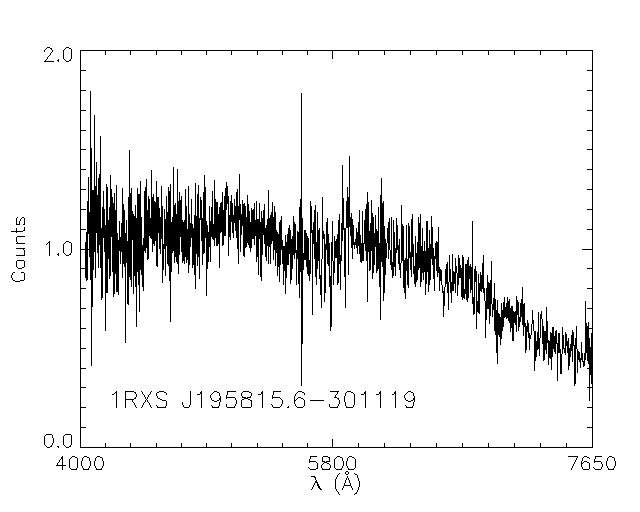} \includegraphics[width=0.45\hsize,height=0.20\hsize]{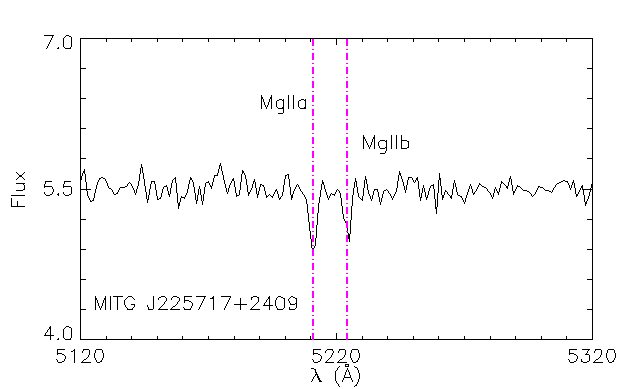}
\includegraphics[width=0.45\hsize,height=0.20\hsize]{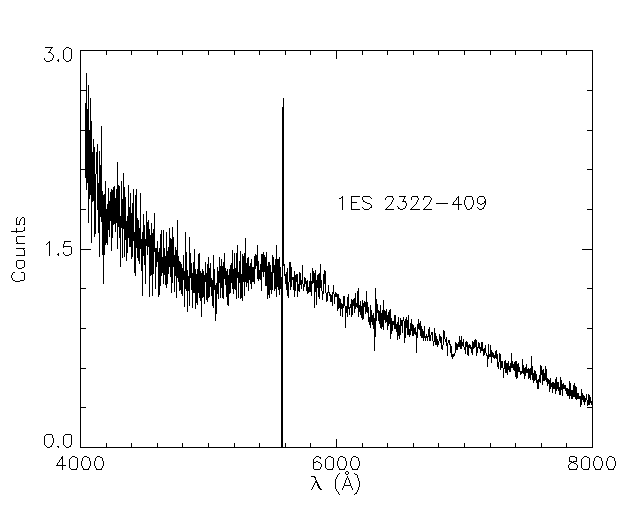}

\caption{Public SDSS and 6dF spectra of sources whose redshift values in our sample are different to those in the 3FHL catalogue. When needed, the lines used to measure the redshift are marked in the plots. First row, left: 1RXS J0022$+$0006, $z$ = 0.3057; right: PMN J0622$-$2605, redshift unknown. Second row, left: NVSS J110735$+$022225, $z \ge$  1.0743; right: S4 1250$+$53, $z \ge$ 0.664. Third row, left: NVSS J141028+143841, $z$ = 0.144; right: SUMSS J144236$-$462302, $z$ = 0.1026. Fourth row, left: RBS 1432, $z \ge$ 0.739; right: 1RXS J195815.6$-$301119, redshift unknown (but $z$ = 0.119 confirmed here; see section 6.15). Fifth row, left: MITG J225717$+$2409, $z \ge$ 0.864; right:  1ES 2322$-$409, redshift unknown (but $z$ = 0.1738 measured here; see section 6.19). }
\label{fig_newplots}
\end{center}
\end{figure*}

\begin{itemize} 
\item {\bf 1RXS J002200.9$+$000659}: This source has no redshift in the 3FHL catalogue, but its spectrum from the SDSS database (Fig. \ref{fig_newplots}, first row, left) clearly shows the features of an elliptical galaxy at $z$ = 0.3057. This is the value that we adopted in our simulations.

\item {\bf KUV 00311$-$1938}: This source has redshift $z$=0.61 \citep{Pir07} in the 3FHL catalogue. Several other high S/N spectra \citep[see e.g.] [] {Pit14} did not show any features except for an MgII absorption system at $z$ = 0.506. We thus adopted $z$ = 0.506  for this source in our simulations.

\item {\bf  PKS 0118$-$272}: The redshift of PKS 0118$-$272 is unknown in the 3FHL catalogue, but several authors \citep[see e.g.] [] {Vla97} have detected a rich absorption system at $z$ = 0.558. This is the value that we adopted in our simulations for the redshift of the source.

\item {\bf MG1 J021114$+$1051}: This source has $z$=0.2 \citep{Mei10} in the 3FHL catalogue. This value is obtained from statistical estimations using imaging observations  that we do not consider accurate enough for our purposes. We therefore consider its redshift unknown.

\item {\bf RBS 334}: This source has no redshift in the 3FHL catalogue. Its redshift has been measured to be $z$ = 0.411 in \citet{Pit14}  and we used this value in our simulations.

\item {\bf NVSS J033859$-$284619}: The redshift of NVSS J033859$-$284619 is unknown in the 3FHL catalogue. \citet{Hal97} determined that its redshift is $z$ = 0.251. This is the value that we used in our simulations

\item  {\bf PKS 0447$-$439}: This source has $z$ = 0.205 \citep{Per98} in the 3FHL catalogue. This value was obtained from the detection of weak features, but it has never been independently confirmed with optical spectroscopy. Furthermore, higher S/N spectra \citep[see e.g.] [] {Pit14} obtained later failed to confirm it, we thus consider the redshift of PKS 0447$-$439 to be unknown.

\item  {\bf 1ES 0502$+$675}:  This source has $z$=0.416 \citep{Landt02} in the 3FHL catalogue. A value of $z$ = 0.314 has also been reported but not discussed in \citet{Sca99}. This puzzling situation may be linked to a contaminating object, identified as a star \citep{Gio04}, which lies at just 0.33 arcsec from the BL Lac object \citep{Sca99}. This distance is smaller than atmospheric seeing, which implies that in every ground observation the spectrum of the BL Lac object is contaminated by the features of the star, which should be subtracted before attempting to measure the BL Lac redshift. We conclude therefore that there is no convincing, precise measurement of the redshift of 1ES 0502$+$675, and we consider it unknown.

\item {\bf TXS 0518$+$211}: This source has $z$ = 0.108 \citep{Shaw13} in the 3FHL catalogue obtained from weak emission features. This value has not been confirmed independently with spectroscopy. Furthermore, higher S/N spectra \citep[see e.g.] []{Pai17a} obtained later failed to confirm this result. We thus conclude that the redshift of TXS 0518$+$211 is unknown.

\item {\bf PKS 0548$-$322}: The redshift of PKS 0548$-$322 is unknown in the 3FHL catalogue, but its spectrum corresponds to that of an elliptical galaxy at $z$ = 0.069 \citep[see e.g.] [] {Fos76} and we took this value as its redshift.

\item {\bf B3 0609$+$413}: This source has no redshift in the 3FHL catalogue, but an MgII system at $z$ = 1.108 has been detected in his spectrum \citep{Shaw13}. We adopted this value as its redshift.

\item {\bf PMN J0622$-$2605}: The redshift of PMN J0622$-$2605 in the 3FHL catalogue is $z$ = 0.414 obtained from a spectrum in the 6dF database \citep{Jon04, Jon09}. However an  examination of the spectrum (see fig. \ref{fig_newplots}, first row, right) reveals no convincing feature in the 6dF spectrum, and we consider its redshift unknown.

\item {\bf 1ES 0647$+$250}: This source has redshift $z$ = 0.203 in the 3FHL catalogue. This value \citep{Rec03} has been classified as tentative. Recent high S/N spectra \citep{Pai17a} failed to show any features, and therefore we consider its redshift unknown.

\item {\bf PMN J0816$-$1311}: PMN J0816$-$1311 has redshift $z$ = 0.046 in the 3FHL catalogue  obtained from a 6DF spectrum \citep{Jon04, Jon09}.  However, an MgII system at $z$ = 0.288 has been detected in its spectrum with X-Shooter  \citep{Pit14}. Its redshift is thus higher than 0.288, and we adopted $z$ = 0.3 as its redshift in our simulations.

\item {\bf NVSS J110735$+$022225}: The redshift of  NVSS J110735$+$022225 is unknown in the 3FHL catalogue. An MgII system at $z$ = 1.0743 has been detected in its SDSS spectrum (Fig. \ref{fig_newplots}, second row, left), and we took this value as its redshift.

\item {\bf RBS 0970}: This source has redshift $z$ = 0.124 in the 3FHL catalogue. This value \citep{Per96} has not been confirmed in several subsequent observations \citep[see e.g.] [] {Pai17b}. We therefore consider its redshift unknown.

\item {\bf S4 1250$+$53}: The redshift of S4 1250$+$53 is unknown in the 3FHL catalogue. An MgII system at $z$ = 0.664 has been detected in its SDSS spectrum (\ref{fig_newplots}, second row, right) and we adopted this value as its redshift.

\item {\bf NVSS J141028$+$143841}: This source has no redshift in the 3FHL catalogue, but its SDSS spectrum shows the typical features of an elliptical galaxy at $z$ = 0.144 (Fig. \ref{fig_newplots}, third row, left), which we took as its redshift.

\item {\bf PKS 1424$+$240}: The redshift of PKS 1424$+$240 is unknown in the 3FHL catalogue. A hydrogen Ly-$\alpha$ absorption system at $z$ = 0.604 has been detected in its spectrum \citep{Fur13}, and we adopted this value as its redshift in our simulations.

\item {\bf RBS 1409}: This source has redshift $z$ = 0.15 \citep{Schwope00} in the 3FHL catalogue, which has not been confirmed in subsequent observations \citep{Shaw13}. We therefore consider its redshift unknown.

\item {\bf SUMSS J144236$-$462302}: This source has no redshift in the 3FHL catalogue, but its SDSS spectrum is that of an elliptical galaxy at $z$ = 0.1026 (Fig. \ref{fig_newplots}, third row, right), which we took as its redshift.

\item {\bf PKS 1440$-$389}: The redshift of PKS 1440$-$389 in the 3FHL catalogue is $z$ = 0.069, obtained from a low S/N spectrum in the 6dF database  \citep{Jon04, Jon09}. Subsequent high S/N spectra failed to show any features, and we considered its redshift as unknown in our simulations. In this paper we report that its redshift is $z$ = 0.1385 (see Section 6.4 for details).

\item {\bf RBS 1432}: The redshift of RBS 1432 is unknown in the 3FHL catalogue. An MgII system at $z$ = 0.739 has been detected in its SDSS spectrum (Fig. \ref{fig_newplots}, fourth row, left), and we took this value as its redshift.

\item {\bf PMN J1603$-$4904}: This source has no redshift in 3FHL. Its redshift has been measured as $z$ = 0.232 \citep{Gol16}, and we used this value in our simulations.

\item {\bf PMN J1610$-$6649}: The redshift of PMN J1610$-$6649 is unknown in the 3FHL catalogue. An MgII system at $z$ = 0.447 has been detected in its spectrum \citep{Shaw13}, and we took this value as its redshift.

\item {\bf TXS 1902$+$556}: Same as the source above, the MgII system is at $z$ = 0.727.

\item  {\bf PMN J1918$-$4111}: Same as the source above, the MgII system is at $z$ = 1.591.

\item  {\bf RX J1931.1$+$0937}:  Same as the source above, the MgII system is at $z$ = 0.476.

\item  {\bf 1RXS J195815.6$-$301119}: 1RXS J195815.6$-$301119 has a redshift of $z$ = 0.119 in the 3FHL catalogue from a low S/N spectrum in the 6dF database (Fig. \ref{fig_newplots}, fourth row, right). Its redshift value was considered unknown in our simulations, but we confirmed it at $z$ = 0.119 in this paper (see Section 6.15 for details).

\item {\bf RGB J2243$+$203}: This source has no redshift in the 3FHL catalogue. An MgII system at $z$ = 0.395 has been detected in its spectrum \citep{Shaw13}, and we took this value as its redshift.

\item  {\bf MITG J225517$+$2409}:  The redshift of MITG J225517$+$2409 is unknown in the 3FHL catalogue. An MgII system at $z$ = 0.864 has been detected in its SDSS spectrum (Fig. \ref{fig_newplots}, fifth row, left), and we took this value as its redshift.

\item  {\bf 1ES 2322$-$409}: 1ES 2322$-$409 has redshift $z$ = 0.174 in the 3FHL catalogue from a low S/N spectrum in the 6dF database (Fig. \ref{fig_newplots}, fifth row, right). Its redshift value was considered unknown in our simulations, but we measured it at $z$ = 0.1738 in this paper (see Section 6.19 for details).
\end{itemize} 

\clearpage

\clearpage

\section{{\em Swift}-UVOT and REM/ROSS photometry}

 \begin{figure*}[!h]
   \centering
  \includegraphics[width=0.5\hsize]{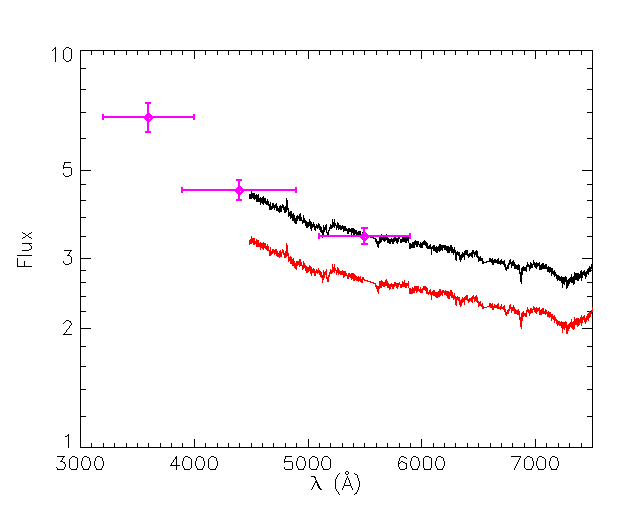}\includegraphics[width=0.5\hsize]{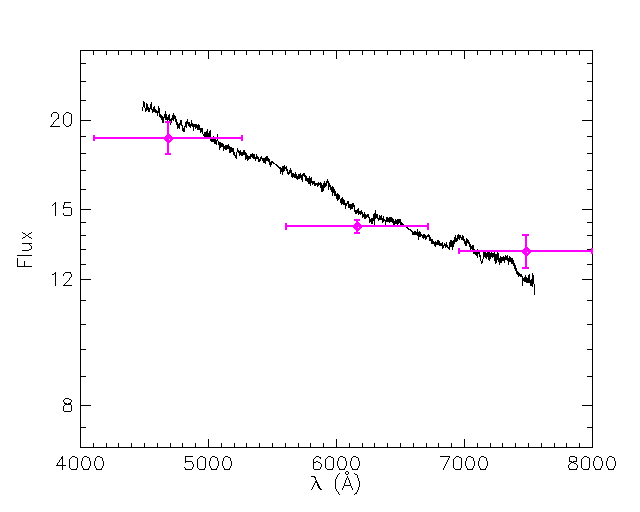}
   \caption{
 Spectra and near-contemporaneous photometry of 1RXSJ015658.6$-$530208 and 1RXS J020922.2-522920. Left panel: SALT/RSS flux calibrated and telluric corrected optical spectrum of 1RXSJ015658.6$-$530208 (in red), and the same spectrum scaled to the average UVOT photometry (in black). The $u$, $b$, and $v$ photometry and errors are shown in magenta. Right panel: SALT/RSS flux calibrated and telluric corrected optical spectrum of 1RXS J020922.2-522920. The $g$, $r$, and $i$ average photometry is shown in magenta. The photometric points are consistent with the spectral flux within  2$\sigma$ at most.}
 
 \label{SALT_photo}
    \end{figure*}

\begin{table*}[!h]
\caption{ Log and results of Swift/UVOT observations of 1RXS J015658.6$-$530208 in $v$, $b$, and $u$ bands.}
\label{tabUVOT}
\begin{center}
\begin{tabular}{ccccc}
\hline
\multicolumn{1}{c}{ Date (UT)}  &
\multicolumn{1}{c}{ MJD}  &
\multicolumn{1}{c}{ $u$ } &  
\multicolumn{1}{c}{$b$ } &
\multicolumn{1}{c}{ $v$ } \\
\multicolumn{1}{c}{} &
\multicolumn{1}{c}{} &
\multicolumn{1}{c}{ (mag) } &
\multicolumn{1}{c}{(mag) } &
\multicolumn{1}{c}{(mag)} \\
\hline
 2019-11-30  & 58817  & 17.12 $\pm$ 0.05  & 18.11 $\pm$  0.06  &   17.72 $\pm$ 0.09   \\
\hline
\end{tabular}
\end{center}
\end{table*}

\begin{table*}[!h]
\caption{ Log and fitting results of REM observations of 1RXS J020922.2$-$522920 in $g$, $r$, and $i$ bands.}
\label{tabREM2}
\begin{center}
\begin{tabular}{ccccc}
\hline
\multicolumn{1}{c}{ Date (UT) } &
\multicolumn{1}{c}{ MJD } &
\multicolumn{1}{c}{ $g$ } &  
\multicolumn{1}{c}{ $r$ } &
\multicolumn{1}{c}{ $i$ } \\
\multicolumn{1}{c}{} &
\multicolumn{1}{c}{} &
\multicolumn{1}{c}{ (mag) } &
\multicolumn{1}{c}{ (mag) } &
\multicolumn{1}{c}{ (mag) } \\
\hline
2019-12-25 &  58842 &  15.839 $\pm$ 0.126    & 15.655 $\pm$  0.067   & 15.334 $\pm$ 0.113 \\
2019-12-31  &  58848    &  16.139 $\pm$ 0.025    &  15.780 $\pm$  0.022   & 15.425 $\pm$ 0.084    \\
2020-01-01 & 58849   &  16.123 $\pm$ 0.150  & 15.858 $\pm$  0.013 &  15.522 $\pm$ 0.141   \\
2020-01-02 &  58850 & 16.204 $\pm$ 0.025 &  15.865 $\pm$0.029 & 15.507$\pm$ 0.053   \\
2020-01-03 &  58851  &  16.218 $\pm$ 0.026 & 15.842 $\pm$  0.028  & 15.491 $\pm$ 0.060   \\
\hline
\end{tabular}
\end{center}
\end{table*}

\end{appendix}

%
%

\end{document}